\documentclass[11pt]{article}
\usepackage{amsmath}
\usepackage{amssymb}
\usepackage{url}
\usepackage{times}
\usepackage{latexsym}
\usepackage{graphicx}
\usepackage{amsthm}
\usepackage{graphicx}
\usepackage{epsfig}
\usepackage{color}
\usepackage{lscape}
\input{Qcircuit.tex}

\newcommand{\be}{\begin{equation}}
\newcommand{\ee}{\end{equation}}
\newcommand{\ba}{\begin{array}}
\newcommand{\ea}{\end{array}}
\newcommand{\bea}{\begin{eqnarray}}
\newcommand{\eea}{\end{eqnarray}}

\newcommand{\dat}[3]{$(#1\pm #2)\times 10^{-#3}$}

\begin{document}

\title{A Comparative Code Study for Quantum Fault Tolerance}

\author{Andrew Cross \thanks{Department of Electrical Engineering and
Computer Science, MIT, Cambridge, MA, 02139, USA and
IBM Watson Research Center, P.O. Box 218, Yorktown Heights, NY, USA
10598. \texttt{andrew.w.cross@gmail.com}} \and David P. DiVincenzo
\thanks{IBM Watson Research Center, P.O. Box 218, Yorktown Heights,
NY, USA \texttt{divince@watson.ibm.com}} \and Barbara M. Terhal
\thanks{IBM Watson Research Center, P.O. Box 218, Yorktown Heights,
NY, USA 10598. \texttt{bterhal@gmail.com}}}

\maketitle
\begin{abstract}
We study a comprehensive list of quantum codes as candidates for
codes used at the physical level in a fault-tolerant code
architecture. Using the Aliferis-Gottesman-Preskill (AGP) ex-Rec
method we calculate the pseudo-threshold for these codes against
depolarizing noise at various levels of overhead. We estimate the
logical noise rate as a function of overhead at a physical error
rate of $p_0=1 \times 10^{-4}$. The Bacon-Shor codes and the Golay
code are the best performers in our study.
\end{abstract}

\tableofcontents


\section{Introduction}



A great insight in the early history of quantum computing was that
almost perfectly reliable quantum computation is possible with
physical devices subjected to noise as long as the noise level is
not too large. This observation has given us confidence that,
ultimately, it will be possible to build a functioning quantum
computer. However, the ``threshold theorem" that indicates how
much noise can be tolerated has not otherwise given a very
optimistic prognosis for progress. For example, the well-studied
seven qubit Steane code [[7,1,3]] has a threshold against
adversarial noise that is in the $O(10^{-5})$ range. This level of
noise is far lower than anything that has been achieved in any
laboratory -- it is actually a significantly lower error rate than
many experiments are even capable of measuring.

In this paper we analyze both previously considered and new
additional quantum codes and determine their thresholds, logical
error rates and overheads using the ex-Rec method developed in
\cite{AGP:ft}.
This allows us for the first time to compare the relative merit of
many schemes. We will argue how the studied codes could serve as
{\em bottom} (physical-level) codes in a fault-tolerant code
architecture that minimizes the overall coding overhead.

In order to carry out this comparative code study we must make some
simplifying assumptions. First of all, we assume that gates can be
performed in a non-local geometry. It is likely that an ultimate
quantum architecture will be largely restricted to local 1D, 2D or 3D
geometries, hence the threshold numbers that we estimate will be
affected by this architecture constraint. As was shown in
\cite{STD:local,SR:localft}, the price for locality may be modest
for small codes, but it will typically be worse for larger codes
\cite{STD:local} since a bigger effort will have to be mounted to
make the error-correction circuits local.

Secondly, the noise model for our study is a simple depolarizing
noise model with equal probabilities for Pauli $X,Y$ and $Z$ errors.
In particular, we assume that any single qubit location in the
quantum circuit undergoes a Pauli $X$,$Y$ or $Z$ error with equal
probability $p_0/3$. A two-qubit gate undergoes no error with
probability $1-p_0$ and one of 15 errors with probability $p_0/15$.
The reason for choosing this model is that it is the simplest
unbiased choice for a noise model given that a comparative study of
the performance of codes for general noise models (such as
superoperator noise) is infeasible computationally. We do not expect
the performance of these codes to differ greatly if we would choose
a biased depolarizing noise model, since there are no intrinsic
biases in the codes themselves. To deal with biased noise, it may be
more advantageous to use specific biased code constructions such a
non-square Bacon-Shor or surface code or the use of the repetition
code against high-rate dephasing noise in \cite{AP:biased}.

We do not establish rigorous threshold lower-bounds by counting the
number of malignant sets of faults as in \cite{AGP:ft}. Instead of
counting or sampling of malignant faults, we simply simulate the
depolarizing noise and keep track of when it leads to failures. We
put error bars on our results such that a rigorous lower bound on
the pseudo-threshold is within this statistical error bar. The
level-1 pseudo-threshold is the value of $p_0$ where the level-1
encoded error rate $p_1=p_0$ \cite{SCCA:pseudo}. It is only the
pseudo-threshold that is of interest in this study of bottom codes
since we envision, see Section \ref{sec:coa}, that a different code
would be used in the next level of encoding.

Our study is in some sense a continuation of the first comparative
code study by Steane \cite{steane:overhead}.
Our analysis goes beyond these previous Monte Carlo studies of
quantum fault-tolerance in that it includes more codes and focuses
on a fault-tolerance analysis of the logical CNOT gate. One of the
problems with comparing threshold estimates in the literature is
that threshold numbers for different codes have been obtained by
different methods, some more rigorous than other's. We believe that
it would be advantageous to stick to one clear, rigorously motivated
method. The AGP method has the advantage of being tied to a fully
rigorous analysis \cite{AGP:ft} and the AGP method when used to
(approximately) count the number of malignant sets of faults can
give tight estimates of the (pseudo)-threshold.

With a few exceptions, we use standard Steane error-correction
circuits (and sometimes omit possible code-specific optimizations)
that allows us to compare codes directly. We discuss these possible
code optimizations and our choices in Section \ref{sec:cc}. We do
not separately analyze Knill's post-selected and Fibonacci schemes
\cite{aliferis:fib, knill:nature} but we plot Knill's numbers in
Figure \ref{fig:errorrate}. One scheme that is not included is the
surface code scheme described in Ref. \cite{RHG:topo}.
However, we do study surface codes in the original setting of
\cite{dennis+:top} using Shor error-correction circuits.

\subsection{The Code Architecture}
\label{sec:coa}

The usefulness of error correction in computation is roughly
measured by two parameters. First is the reduction in error rate
that is obtained by using the code; this feature depends on the
(pseudo)-threshold of the code for the particular noise model and
the error correction circuits that are used. The second figure of
merit is the smallness of the overhead that is incurred by coding.
There is a tradeoff that one can expect between overhead and logical
error-rate that mimics the trade-off between distance and rate of
error-correction codes, see Figure \ref{fig:errorrate}. Since error
levels of physical implementations are expected to be high,
optimistically in the range from $O(10^{-2})$ to $O(10^{-6})$, it is
clear that at the bottom level, optimizing the threshold has
priority over optimizing overhead. This leads us to consider the
following simple code architecture. At the physical level, we use a
bottom code $C_{\rm bot}$ which is chosen to have a high noise
threshold and a reasonable overhead. This paper will be devoted to a
comparative study of such codes. We will pick some illustrative
numbers to argue how one can envision completing the code
architecture. We will see that one can find a bottom code that maps
a base error rate of $p_0=O(10^{-4})$ onto a logical error rate of
$p_1=O(10^{-7})$ (see Section \ref{sec:results}). To run a
reasonable-sized factoring algorithm one may need an logical error
rate of, say, $O(10^{-15})$ \footnote{An $n$-bit number can be
factored using a circuit with space-time complexity of roughly
$360n^4$ \cite{beckman:factor}, so RSA-1024 could be broken using a
circuit with $O(10^{15})$ potential fault locations. Using different
architectures, it may be possible to reduce this to $O(10^{11})$ see
for example \cite{vanmeter:architecture}.}. Thus one needs a top
code $C_{\rm top}$ that brings the error rate $O(10^{-7})$ to
$O(10^{-15})$. The desirable features of the top code are roughly as
follows (see also the Discussion at the end of the paper). The top
code is a block code $[[n,k,d]]$ with good rate $k/n$ in order to
minimize the overhead. The improvement in error-rate for a code
which can correct $t$ errors is roughly \be p_1 \approx p_0
\left(\frac{p_0}{p_{\rm th}}\right)^{t}, \label{eq:p1estim} \ee
where $p_0$ is the unencoded error-rate and $p_{\rm th}$ is the
threshold error rate. Thus in order to get from $p_0=O(10^{-7})$ to
$p_1=O(10^{-15})$ we could use a code which can correct $4$ errors
and has a threshold of $O(10^{-5})$.
In \cite{steane:nature} Steane studied several block-codes which may
meet these demands. The polynomial codes
would also be an interesting family to study in this respect.

\section{Preliminaries}
\label{sec:cc}

For our study it is necessary to select a subset of quantum codes.
We focus on codes that are likely to have a good threshold,
possibly at the cost of a sizeable but not gigantic overhead. To
first approximation the threshold is determined by the equation
\be p_{\rm th}= N p_{\rm th}^{t+1} \Rightarrow p_{\rm
th}=N^{-1/t}, \label{eq:thresh} \ee where $t$ is the number of
errors that the code can correct and $N$ is a combinatorial factor
counting the sets of $t+1$ locations in an encoded gate that lead
to the encoded gate failing. It is clearly desirable to minimize
the number of locations and maximize $t$. This consideration has
led us to primarily consider Calderbank-Shor-Steane (CSS) codes.
Any stabilizer quantum code is CSS if and only if the CNOT gate is
a transversal gate \cite{thesis:gottesman}. The advantage of a
transversal CNOT is that it minimizes the size of the encoded
CNOT; the bulk of the CNOT rectangle will be taken up by error
correction. This is favorable for the noise threshold of $C_{\rm
bot}$. Secondly, minimizing the error-rate of the encoded gate
$C_{\rm bot}({\rm CNOT})$ will be useful at the next level of
encoding, because CNOTs occur frequently in EC and their error
rates play a large role in determining whether error rates are
below the threshold (of $C_{\rm top}$).
However to demonstrate that this restriction to CSS codes is
warranted we also consider the non-CSS 5-qubit code $[[5,1,3]]$
which is the smallest code that can correct a single error. We
indeed find that this code performs {\em worse} than Steane's
7-qubit code $[[7,1,3]]$, see Section \ref{sec:results} and the Data
Tables in Appendix \ref{app:data}.

\subsection{Approximate Threshold Scaling}
\label{sec:scaling}

In this section we discuss the global behavior of the noise
threshold as a function of block size $n$, distance, and other code
properties. Let us consider Eq. (\ref{eq:thresh}) and see how we can
get the best possible threshold. An upper-bound on $N$ is ${A
\choose t+1}$ where $A$ is the total number of locations in the
encoded gate (rectangle). Ideally, a code family has a distance that
is linear in $n$, i.e. $t$ is linear in $n$. Let us assume for
simplicity that only some fraction of all locations appears in the
malignant fault sets of size $t+1$, i.e. we model $N \approx {A_{\rm
mal} \choose t+1}$ where $A_{\rm mal} < A$. The locations in $A_{\rm
mal}$ are in some sense the weak spots in the circuits; overall
failure is most sensitive to failure at these locations. $A_{\rm
mal}$ may be either linear or super-linear in the block size $n$. In
case $A_{\rm mal}$ scales linearly with $n$, {\em and} $t=\delta n$
for some $\delta \leq 1/4$, the threshold in Eq. (\ref{eq:thresh})
{\em increases} as a function of $n$ and asymptotes in the limit of
large $n$ to a finite value. Indeed, for $A_{\rm mal}=\alpha n$ and
$\delta \ll \alpha$ (which is typically the case since $t \leq n/4$)
we get, using Sterling's approximation,
\begin{equation}
p_{\rm th}=\lim_{n\rightarrow\infty}{\alpha n \choose \delta
n+1}^{-1/(\delta n)}=\frac{\delta}{e \alpha}+
O\left(\frac{\delta^2}{\alpha^2}\right)
\end{equation}
Such monotonic increase of the threshold with block-size is clearly
desirable. It is also clear that when $t$ is constant, for any
polynomial $A_{\rm mal}={\rm poly}(n)$, the threshold $p_{\rm th}$
in Eq. (\ref{eq:thresh}) decreases as a function of $n$. When
$A_{\rm mal}$ scales super-linearly with $n$ and $t$ is linear in
$n$ we get the following behavior. First, the threshold increases
with $n$ (the effect of larger $t$), then the threshold declines
since the effect of a super-linear $A_{\rm mal}$ starts to dominate.
For codes and EC circuits with this behavior, it is thus of interest
to determine where this peak threshold performance occurs. We will
see some evidence of these peaks in Figure \ref{fig:block} in
Section \ref{sec:results}.

Now let us consider the scaling of $A$ (and $A_{\rm mal}$) in case
we use Steane error correction. In Appendix \ref{app:css} we
review how we can bound $A$ for a CSS code with Steane error
correction, but a rough estimate is that \be A=c_1 A_{\rm enc}+
c_2 A_{\rm ver}+c_3 n. \ee Here $A_{\rm enc}$ is the number of
locations in the encoding of the ancillas for error correction,
and $A_{\rm ver}$ is the number of locations in the verification
of the ancillas for error correction. The additional term linear
in $n$ comes from the transversal encoded gates and the
transversal syndrome extractions. For a CSS code and the standard
encoding construction (see Appendix \ref{app:css}), $A_{\rm enc}$
typically scales as $O(w n)$ where $w$ is the maximum Hamming
weight of the rows of the generator matrix of either $C_1$ or
$C_2^{\perp}$ {\em in standard form}. However this standard
construction may be sub-optimal, since by bringing the generator
matrix in standard form one can increase the maximum weight of its
rows.


For Steane-EC the full verification of the ancilla block requires
other ancillas blocks; a fully fault-tolerant verification would
give a pessimistic scaling of $A_{\rm ver}=O(w n t)$. However it is
not necessarily desirable to have strict fault tolerance as long as
the total probability of low-weight faults that produce faults with
weight $t+1$ or more is low, see the discussion in Section
\ref{sec:methods}. On the other hand for increasing $n$ the number
of verification rounds should at least be increasing with $n$,
perhaps $O(\log n)$ would be sufficient. If we assume that $A_{\rm
mal}$ scales similarly as $A$, it follows that if we look for
linear-scaling $A_{\rm mal}$ we need to look at code families which
have simple encoders, scaling linearly with $n$. This seems only
possible for stabilizer codes with constant weight stabilizers, such
as quantum LDPC codes \cite{MMM:ldpc} and surface codes or for the
Bacon-Shor codes (which has encoders that use $O(n)$ 2-qubit gates).

For the Bacon-Shor and surface codes the distance $t$ does not
scale linearly with $n$ (but as $\sqrt{n}$). Nonetheless, the work
in \cite{dennis+:top} shows that the effective distance for the
surface codes does scale linearly with the block size, since there
are very few incorrectible errors of weight $O(t)$. For the
Bacon-Shor code family, where one has less syndrome information,
this behavior has not been observed \cite{AC:bs} (see also Figure
\ref{fig:perfectlonger}).

For code families with constant-weight stabilizers an interesting
alternative to Steane-EC \cite{steane:active} is the use of Shor-EC
\cite{NC:book} where the syndrome corresponding to each stabilizer
is extracted using a cat state or simple unencoded qubit ancillas.
As for ancilla verification in Steane-EC, the syndrome extraction
needs to be repeated to make the circuits more fault-tolerant. In
Section~\ref{sec:results} we will see the effect on the threshold of
using Steane-EC versus Shor-EC for the surface codes
\cite{BK:surface,FM:surface}, see Figure~\ref{fig:surfacethreshold}
of Section~\ref{sec:results}. It is striking that the surface codes
with Shor EC are the only known examples of a code family with a
finite $n \rightarrow \infty$ threshold. This is despite the
$O(n\sqrt{n})$ scaling of the total number of locations $A$ of the
Shor error correction circuit.



\subsection{Choice of Codes}

The codes that we have studied are listed in Table~\ref{tab:codes}.
All codes in this table are CSS codes with the exception of the
[[5,1,3]] code. Some of these codes have been previously analyzed by
Steane in Ref. \cite{steane:overhead}. There exist various families
of binary CSS quantum codes; the families are the quantum
Reed-Muller codes, the quantum Hamming codes, the quantum BCH codes,
the surface codes and the sub-system Bacon-Shor codes. In our study
we consider only a single member of the quantum Reed-Muller family,
a $[[15,1,3]]$ code, since these codes typically don't have very
good distance versus block-size \cite{steane:rm}. The $[[15,1,3]]$
code was first constructed in \cite{KLZ:faulttol} from a punctured
Reed-Muller code ${\rm RM}(1,4)$ and its even sub-code. It is the
smallest known distance-$3$ code with a transversal T gate.

We study various quantum Hamming and quantum BCH codes (see a
complete list of quantum BCH codes of small block-size in
\cite{grassl:bchcodes}) which are constructed from self-orthogonal
classical Hamming and BCH codes respectively. We have chosen those
codes that encode a single qubit and have maximum distance for a
given block size. We have included the previously studied
Bacon-Shor codes and the surface codes in our study. We have also
included the concatenated 7-qubit code $[[49,1,9]]$ which we use
in the way that was proposed by Reichardt in
\cite{reichardt:concat7}, see the details in Section
\ref{sec:FTECspecial}.

Another family of codes that has been proposed for fault tolerance
\cite{AB:ftsiam} are the quantum Reed-Solomon codes or polynomial
codes. These are codes that are naturally defined on qudits. In
this study we consider them as candidates for bottom codes. An
alternative use is to consider them as top codes where one uses a
bottom code to map the qubits onto qudits. In our study we assume
that quantum information is presented in the form of qubits and
hence we will consider these codes as binary stabilizer codes.
We specifically chose to include the $[[21,3,5]]$ (a concatenated
$[[7,1,4]]_8$) and the $[[60,4,10]]$ (a concatenated
$[[15,1,8]]_{16}$) code from the family of dual-containing
polynomial codes over $GF(2^m)$, because they are the smallest
error-correcting polynomial codes in this family.

We find it impractical to simulate the encoded CNOT gate for BCH
codes in this table which have block-size larger than $[[47,1,11]]$,
see \ref{sec:software}. The threshold for these bigger codes will
benefit considerably from the fact that $t/n$ is quite high. Some
semi-analytical values for the thresholds of these codes have been
given in \cite{steane:overhead}. Even with good thresholds, these
bigger BCH codes have limited applicability due to their large
overhead. The bottom code should be picked to obtain a logical error
rate that is well below the threshold of some good block code but
only at the price of a moderate overhead.


\begin{table}
\begin{center}
\begin{tabular}{l|l|l}
\textsc{Parameters}  & \textsc{Notes}
\\ \hline
$[[5,1,3]]$ & non-CSS five qubit code \cite{laflamme+:5qubit}\\
$[[7,1,3]]$ & Steane's 7-qubit code (doubly-even dual-containing) \cite{steane:7qubit} \\
$[[9,1,3]],[[25,1,5]],[[49,1,7]],[[81,1,9]]$ &  Bacon-Shor codes \cite{AC:bs} \\
$[[15,1,3]]$ & Quantum Reed-Muller code \cite{steane:rm,KLZ:faulttol} \\
$[[13,1,3]],[[41,1,5]],[[85,1,7]]$ & Surface codes \cite{BK:surface,FM:surface}\\
$[[21,3,5]]$ & Dual-containing polynomial code on $GF(2^3)$ \cite{grassl+:rscode} \\
$[[23,1,7]]$ & Doubly-even dual-containing Golay code (cyclic) \cite{thesis:reichardt} \\
$[[47,1,11]]$ & Doubly-even dual-containing quadratic-residue code (cyclic) \cite{grassl:bchcodes} \\
$[[49,1,9]]$ & Concatenated $[[7,1,3]]$ Hamming code \cite{reichardt:concat7} \\
$[[60,4,10]]$ & Dual-containing polynomial code on $GF(2^4)$ \cite{grassl+:rscode} \\
$[[79,1,15]],[[89,1,17]],[[103,1,19]],[[127,1,19]]$ & BCH codes, not analyzed \cite{grassl:bchcodes} \\
\end{tabular}
\end{center}
\caption{A list of the codes included in our study.} \label{tab:codes}
\end{table}

\subsection{Universality}
\label{sec:univ}

Universality for CSS codes can in principle be obtained using the
technique of {\em injection-and-distillation}
\cite{knill:nature,reichardt:distill, BK:magicdistill}. Let us
briefly review how one may perform fault-tolerant computation for
CSS codes for which, of the Clifford group gates, only the CNOT and
Pauli operations are transversal. If one is able to perform any
Clifford group gate transversally, including H and S, it is well
known how to obtain a universal set of gates \cite{thesis:aliferis}.
Note that a CSS code with only its transversal CNOT gives us the
ability to fault-tolerantly prepare the states
$\{|\overline{+}\rangle,|\overline{0}\rangle\}$ and perform
transversal $\overline{X}$ and $\overline{Z}$ measurements. However
we do not necessarily have a fault-tolerant realization of the
Hadamard gate H.

\begin{figure}[htb]
\centerline{
\mbox{ \Qcircuit @C=1em @R=.5em {
              & \ctrl{1}     & \measuretab{\mbox{X-basis}} & \control \cw \cwx[1] \\
\lstick{\ket{\textrm{+}i}} & \targ & \qw            & \gate{Y} & \qw
}} } \caption{The Q gate using a $|\textrm{+}i\rangle$ ancilla.}
\label{fig:Qgate}
\end{figure}
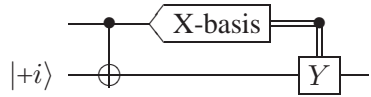

In this case the problem of constructing fault-tolerant single
qubit Clifford gates can be reduced to the problem of preparing
the encoded $|\overline{\textrm{+}i}\rangle\propto
|\overline{0}\rangle+i|\overline{1}\rangle$ ancilla
\cite{preskill_aliferis:comm}. In particular, the gates
${\rm S}\propto\exp(-i\pi Z/4)$ and ${\rm Q}\propto\exp(+i\pi
X/4)$ generate the single-qubit Clifford group and can be
implemented given a $|\textrm{+}i\rangle$ ancilla, see
Figure~\ref{fig:Sgate} and Figure~\ref{fig:Qgate}.

\begin{figure}[htb]
\centerline{
\mbox{ \Qcircuit @C=1em @R=.5em {
              & \targ     & \measuretab{\mbox{Z-basis}} & \control \cw \cwx[1] \\
\lstick{\ket{+i}} & \ctrl{-1} & \qw            & \gate{Y} & \qw }} }
\caption{The S gate using a $|\textrm{+}i\rangle$ ancilla.}
\label{fig:Sgate}
\end{figure}

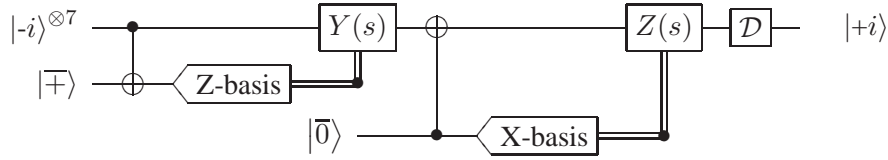
\begin{figure}
\centerline{ \mbox{ \Qcircuit @C=1em @R=.5em {
\lstick{\ket{\textrm{-}i}^{\otimes 7}} & \ctrl{1}     & \qw            & \gate{Y(s)}            & \targ & \qw            & \gate{Z(s)} & \gate{\cal{D}} & \qw & \rstick{\ket{\textrm{+}i}} \\
\lstick{\ket{\overline{+}}}                         & \targ & \measuretab{\mbox{Z-basis}} & \control\cw\cwx[-1]  \\
                                               &           &                & \lstick{\ket{\overline{0}}} & \ctrl{-2}    & \measuretab{\mbox{X-basis}} & \control \cw \cwx[-2]
}} } \caption{A circuit for the distillation of a
$|\textrm{+}i\rangle$ ancilla from seven noisy $\ket{\textrm{-}i}$
ancilla's using the [[7,1,3]] Steane code. Steane error-correction
(see also Figure \ref{fig:SteFTEC}) is performed on seven
unencoded qubits $\ket{-i}$. Based on the $X$ and $Z$ syndromes of
the Steane code the errors are corrected, except that the $X$ errors
are corrected using $Y$ operators. The last step is to decode the
Steane code to yield a single $|\textrm{+}i\rangle$ ancilla. The
circuit can be found by starting from a deterministic distillation procedure
for $|+\rangle$, applying an S gate to the output, and conjugating
it back to the input. The order of error corrections is important since
the second CNOT must be replaced by a controlled-Y before the order can be
reversed.
}
\label{fig:Idistill}
\end{figure}

An encoded $|\overline{\textrm{+}i}\rangle$ ancilla can be produced
using the method of injection-and-distillation
\cite{BK:magicdistill,reichardt:distill}. The distillation procedure
for distilling an unencoded $|\textrm{+}i\rangle$ from seven
unencoded $|\textrm{-}i\rangle$ is shown in
Figure~\ref{fig:Idistill}. In order to perform this distillation
procedure in encoded form, one can generate an encoded noisy ancilla
$|\overline{\textrm{-}i}\rangle$ using Knill's idea of injecting a
state in the code. The distillation circuit is then performed in
encoded form.


\section{Error Correction Circuits}
\label{sec:circ}

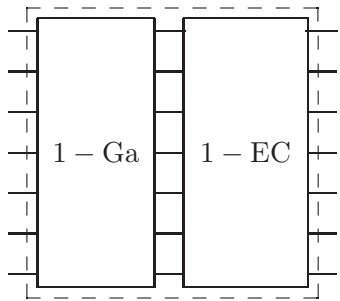
\begin{figure}[htbp]
\centerline{
\mbox{ \Qcircuit @C=1em @R=.5em {
      & \multigate{6}{{\rm 1-Ga}} & \multigate{6}{{\rm 1-EC}} & \qw \\
      & \ghost{1-Ga} & \ghost{1-EC} & \qw \\
      & \ghost{1-Ga} & \ghost{1-EC} & \qw \\
      & \ghost{1-Ga} & \ghost{1-EC} & \qw \\
      & \ghost{1-Ga} & \ghost{1-EC} & \qw \\
      & \ghost{1-Ga} & \ghost{1-EC} & \qw \\
      & \ghost{1-Ga} & \ghost{1-EC} & \qw \gategroup{1}{2}{7}{3}{.7em}{--} }} }
\vspace*{13pt} \caption{\label{fig:1Rec} A 1-rectangle (1-Rec),
indicated by a dashed box, which replaces a single-qubit 0-Ga
location. The 1-Rec consists of the encoded fault-tolerant
implementation of the 0-Ga (1-Ga) followed by an error correction
procedure (1-EC).}
\end{figure}

{\em Locations} in a quantum circuit are defined to be gates,
single-qubit state preparations, measurement steps, or memory (wait)
locations. After one level of encoding, every location (denoted as
0-Ga) is mapped onto a a rectangle or 1-rectangle (1-Rec), a
space-time region in the encoded circuit, which consists of the
encoded gate (1-Ga) followed by error correction (1-EC), as shown in
Fig.~\ref{fig:1Rec}. For transversal gates, the 1-Ga consists of
performing the 0-Ga's on each qubit in the block(s).

For the fault-tolerance analysis one also defines an extended
1-Rec or ex-Rec which consists of a rectangle along with its {\em
preceding} 1-EC(s) on the input block(s). Let us now discuss the
circuits for error correction.


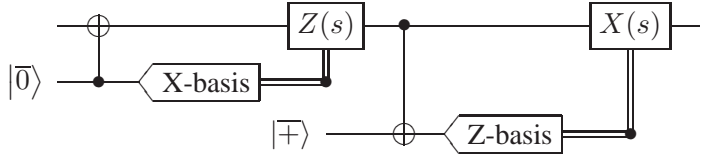
\begin{figure}[htb]
\centerline{
\mbox{ \Qcircuit @C=1em @R=.5em {
                       & \targ     & \qw            & \gate{Z(s)}            & \ctrl{2} & \qw            & \gate{X(s)} & \qw \\
\lstick{\ket{\overline{0}}} & \ctrl{-1} & \measuretab{\mbox{X-basis}} & \control\cw\cwx[-1]  \\
                       &           &                & \lstick{\ket{\overline{+}}} & \targ    & \measuretab{\mbox{Z-basis}} & \control \cw \cwx[-2]
}} } \caption{Steane's error correction method for CSS codes
involves coupling two encoded and verified ancilla's to a block of
data qubits. The ancilla qubits are then measured in the $Z$- or
$X$-basis and the syndrome $s$ is determined. From the syndrome $s$
the corresponding $Z$ or $X$ error is determined and the data qubits
are corrected.} \label{fig:SteFTEC}
\end{figure}

\begin{figure}[htb]
\centerline{
\mbox{ \Qcircuit @C=1em @R=.5em {
                       & \ctrl{1}  & \qw            & \targ     & \qw            \\
\lstick{\ket{\overline{0}}} & \targ     & \measuretab{\mbox{Z-basis}} \\
\lstick{\ket{\overline{0}}} & \ctrl{1}  & \qw            & \ctrl{-2} & \measuretab{\mbox{X-basis}} \\
\lstick{\ket{\overline{0}}} & \targ     & \measuretab{\mbox{Z-basis}} \\
}}} \caption{The ancilla verification circuit for one round of error
detection. First, the top encoded ancilla is verified against
$X$-errors and it is determined whether $\overline{Z}=+1$. If no
$X$-errors are detected and $\overline{Z}=+1$ the encoded ancilla is
then verified against $Z$-errors using an ancilla
$\ket{\overline{0}}$ which is itself verified against $X$-errors.}
\label{fig:singlestage}
\end{figure}
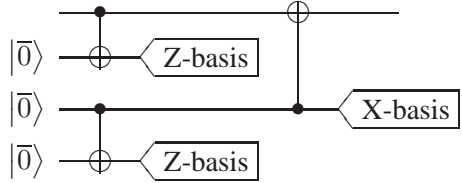

Steane error correction for CSS codes (Steane-EC) is schematically
shown in Figure~\ref{fig:SteFTEC}. The $|\overline{0}\rangle$ and
$|\overline{+}\rangle$ ancilla blocks in Fig. \ref{fig:SteFTEC} can
be prepared in the following way. First $n$ qubits are encoded using
circuits derived from the generator matrix of a classical coset code
of $C_2^\perp$, see Appendix \ref{app:css} for details and
definitions. The memory locations in the encoder are determined
using Steane's Latin rectangle method \cite{steane:fast}, discussed
in more detail in Appendix \ref{app:latin}. Then the ancillas pass
through a verification circuit. This error detection circuit
measures the $X$ and $Z$ stabilizer generators of the encoded state
some number $R$ of times. For a $\ket{\overline{0}}$ ancilla each
round is given by the circuit in Figure~\ref{fig:singlestage}. For
dual-containing codes, the Hadamard-conjugate of the circuit is used
for a $\ket{\overline{+}}$
ancilla. If we detect any errors in any of the $R$ rounds, the
encoded state is rejected. Otherwise the state is accepted and used
for syndrome extraction. We will consider $L$ preparation attempts
per ancilla and in our studies we will vary the parameters $R$ and
$L$, giving rise to different overheads.


\subsection{Specific Code Considerations}
\label{sec:FTECspecial}

Specific properties of a quantum code can often be used to
simplify the error-correcting circuits. This section discusses
each family of codes and the optimizations we have implemented or
the code properties that have been used to modify the EC and 1-Ga
circuits.

In general, we have opted to focus on the error-correcting
properties of the codes rather than the possible simplifications to
the Steane-EC network. One of the reasons for this approach is that
it is not clear whether verification circuits that perform the
minimal number of checks are superior to verification circuits that
perform many thorough tests. Furthermore, changes to the network are
difficult to parameterize and systematically study because there are
many possible choices and few are clearly the best. In addition, we
believe that the overall trends observed in this paper are not
altered by omitting these optimizations.

Reichardt has suggested a generic optimization that uses different
encoders for each logical ancilla in the verification circuit
\cite{thesis:reichardt}. This optimization can reduce the number of
necessary rounds of verification and possibly decrease the
probability of correlated errors at the output of the verification
circuit, conditioned on acceptance. We do not use this optimization
for any of the codes in this study.

The Steane and Golay codes are constructed from perfect classical
codes. Perfect codes have the property that every syndrome locates a
unique error of weight $w\leq t$. As Ref. \cite{AGP:ft} observed,
some parts of the error detection circuit can be removed for a CSS
code constructed from perfect classical codes and the construction
remains strictly fault-tolerant. Again we do not use this
optimization.

For the Bacon-Shor codes we don't use Steane's Latin rectangle
encoding method, but rather the simpler method described in
\cite{AC:bs}. We do use the standard verification method for the
bigger BS codes and not the simpler verification method in
\cite{AC:bs}.


For the surface codes we consider both Steane-EC and Shor-EC to
understand their effects on the threshold. We use Shor-EC using
bare ancillas as in \cite{dennis+:top}. This is fault-tolerant for
surface codes on a $5\times 5$ lattice or larger as long as the
syndrome measurements are repeated enough times. The number of
repeated measurements could in principle be varied, but we choose
to repeat the measurements $\ell$ times for a $\ell\times\ell$
surface code, following \cite{dennis+:top}.

The $[[49,1,9]]$ concatenated Steane code is one of the CSS codes in
our study whose network deviates from the construction described in
the previous section. The preparations of $\ket{\overline{0}}$ and
$\ket{\overline{+}}$ do not include a verification circuit. Instead
each 7-qubit block has an error detection step after each
[[7,1,3]]-encoded logical gate \cite{thesis:reichardt}. A $49$-qubit
ancilla is rejected if any of these error-detections detects an
error. This implies that any single fault will be detected, so the
circuit is fault-tolerant. In fact, any pair of faults is also
detected, so that a third order event is necessary to defeat the
error-detections. This way of using $[[49,1,9]]$ is the one which
Reichardt proposed. Unlike in his simulations we restrict ourselves
to a finite number of ancilla preparation attempts $L$, since we
care about the total overhead.

The polynomial codes that we consider are non-binary codes over
$2^m$-dimensional qudits. We can choose the parameters of these
codes so that when we consider each qudit as a block of $m$ qubits,
the Fourier transform and controlled-SUM gates are implemented by
bitwise application of Hadamard and CNOT, respectively. In this
setting, the code is simply a binary CSS code encoding $m$ qubits
which is constructed from a non-binary dual-containing classical
code by concatenating using a self-dual basis
of $GF(2^m)$. The advantage of such a
construction is that we can decode the syndromes as if they were
vectors over $GF(2^m)$, allowing us to correct more higher-order
errors than we could otherwise correct as a binary code. To use this
advantage, we do not need to change the way we construct the
rectangles at all, only the way we interpret the classical
measurement outcomes.

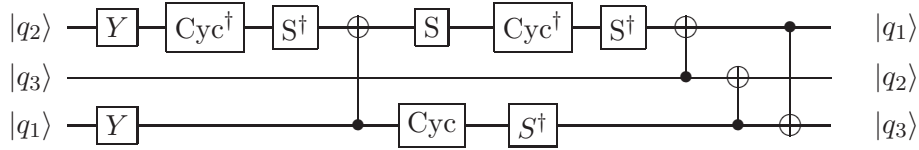
\begin{figure}
\centerline{
\mbox{ \Qcircuit @C=1em @R=.5em {
\lstick{\ket{q_2}} & \gate{Y} & \gate{{\rm Cyc}^\dag} & \gate{{\rm S}^\dag} & \targ     & \gate{{\rm S}}   & \gate{{\rm Cyc}^\dag} & \gate{{\rm S}^\dag} & \targ     & \qw       & \ctrl{2}  & \qw & \rstick{\ket{q_1}} \\
\lstick{\ket{q_3}} & \qw      & \qw             & \qw           & \qw       & \qw        & \qw             & \qw           & \ctrl{-1} & \targ     & \qw       & \qw & \rstick{\ket{q_2}} \\
\lstick{\ket{q_1}} & \gate{Y} & \qw             & \qw           &
\ctrl{-2} & \gate{{\rm Cyc}} & \gate{S^\dag}   & \qw           & \qw
& \ctrl{-1} & \targ     & \qw & \rstick{\ket{q_3}} }} } \caption{The
encoded implementation of ${\rm T_3}$ (with an additional
permutation of the blocks $q_1, q_2, q_3$) using the gates CNOT, S,
Cyc and Y (and inverses). Each gate in the circuit is applied
transversally. The circuit is only a logical operation after
completing all of the gates, i.e. CNOT and S are not valid
transversal gates for $[[5,1,3]]$.}
\label{fig:T3}
\end{figure}

$[[5,1,3]]$ is the smallest distance-$3$ quantum error-correcting
code and it is a perfect quantum code. Gottesman has shown how to
compute fault-tolerantly with this code \cite{thesis:gottesman}, and
there have been some numerical studies of logical error rates using
Shor-EC \cite{fowler:thesis}. To our knowledge the threshold for
this code has never been determined. Unfortunately, there are no
two-qubit transversal gates for $[[5,1,3]]$, so it is necessary to
construct a two-qubit gate from a three-qubit gate such as the ${\rm
T_3}$ gate. The gate ${\rm T_3}$ is defined by the following action
on Pauli operators: $XII \rightarrow XYZ, IXI \rightarrow YXZ, IIX
\rightarrow XXX, ZII \rightarrow ZXY, IZI \rightarrow XZY, IIZ
\rightarrow ZZZ$. This gate is a Clifford gate that can be combined
with stabilizer-state preparations and transversal Pauli
measurements to yield any gate in the Clifford group
\cite{thesis:gottesman}. Specifically, CNOT, S, and Cyc gates (and
their inverses) can be constructed from the ${\rm T_3}$ gate in this
way. Here Cyc$={\rm SHSH}$ acts as $X\rightarrow Y\rightarrow
Z\rightarrow X$. The fault-tolerant implementation of ${\rm T_3}$ is
shown in Figure~\ref{fig:T3}.

The $[[5,1,3]]$ construction also differs from other CSS
constructions because we use Knill (or teleported) error correction
(Knill-EC) \cite{knill:nature}. In our study we will simulate the
extended ${\rm T_3}$-rectangle assuming that the logical Bell pairs
of Knill-EC are perfect. We do this since it is simpler and shows
that even using perfect logical Bell pairs the threshold is not very
good, see Section \ref{sec:results}. For [[5,1,3]] the $R$ and $L$
parameters are replaced by $NC$ and $NB$, denoting the number of cat
state preparation attempts per Pauli measurement and the number of
logical Bell state preparation attempts per error correction,
respectively. A circuit to prepare and verify encoded Bell pairs for
Knill error correction for [[5,1,3]] is shown in
Figure~\ref{fig:bell513} in Appendix \ref{app:decode}.

The construction for the $[[15,1,3]]$ Reed-Muller code is entirely
standard. Since this code is not constructed from a dual-containing
classical code, the $\ket{\overline{0}}$ and $\ket{\overline{+}}$
encoders are not simply related by a transversal Hadamard gate. For
the same reason, the code can correct more $X$ errors than $Z$
errors. The most interesting feature of this code is that T is a
transversal gate \cite{KLZ:faulttol,ZCC:trans}, but this does not
enter directly into our analysis of the threshold.

\section{The Aliferis-Gottesman-Preskill (AGP) Method}
\label{sec:methods}

According to \cite{AGP:ft} a rectangle is {\em correct} if the
rectangle followed by an ideal decoder is equivalent to the ideal
decoder followed by the ideal gate (0-Ga) that the rectangle
simulates:

\begin{center}
\begin{picture}(292,24)
\put(0,12){\line(1,0){10}}
\put(10,0){\framebox(48,24){\shortstack{correct\\$1$-Rec}}}
\put(58,12){\line(1,0){10}}
\put(68,0){\framebox(48,24){\shortstack{ideal\\$1$-decoder}}}
\put(116,12){\line(1,0){10}}
\put(126,6){\makebox(20,12){=}}
\put(146,12){\line(1,0){10}}
\put(156,0){\framebox(48,24){\shortstack{ideal\\$1$-decoder}}}
\put(204,12){\line(1,0){10}}
\put(214,0){\framebox(48,24){\shortstack{ideal\\$0$-Ga}}}
\put(262,12){\line(1,0){10}}
\put(272,6){\makebox(20,12){.}}
\end{picture}
\end{center}

As said before, an {\em extended rectangle} (ex-Rec) consists of a
1-Ga along with its leading and trailing error-corrections. The
extended rectangles make an overlapping covering of the circuit. A
set of locations inside an ex-Rec is called {\em benign} if the
1-Rec is correct for any set of faults occurring on these
locations. If a set of locations is not benign, it is {\em
malignant}. The design principles of {\em strict} fault-tolerance
are described in pictures in Sec. 10 of \cite{AGP:ft}. If these
properties hold for the 1-Ga and 1-EC, these gadgets
for a $[[n,1,d]]$ code with $t=\lfloor(d-1)/2\rfloor$ are called
$t$-strictly fault-tolerant. The important consequence of these
conditions is that for a $[[n,k,d]]$ code with $t$-strictly
fault-tolerant constructions one can show that any set of $t$ or
fewer locations in the ex-Rec is benign. A construction is called
weakly fault-tolerant when, for a code that can correct $t$ errors,
sets of $s < t$ locations can be malignant. Weak fault-tolerance is
a useful concept in optimizing thresholds since weakly
fault-tolerant circuits can be more compact than strictly
fault-tolerant circuits, hence allowing for fewer fault locations
and a potentially better threshold. On the negative side, weak
fault-tolerance allows some low-weight faults to be malignant but if
the number of such faults is small then the threshold is not much
affected.

All our fault-tolerant schemes are $1$-strictly fault-tolerant
implying that single faults can never be malignant. More
precisely, any single fault in a 1-EC or a 1-Ga never propagates
to become a weight-2 error in a block. In Steane-EC when we
prepare ancillas with at least two attempts ($L\geq 2$) and one
error detection stage ($R=1$), we eliminate malignant faults of
weight $1$. For $R=1$ the EC is not 2-strictly fault-tolerant
since there may be a pair of faults, one in each of the first two
encoders, generating a high weight (possibly higher than $t$)
error that passes the error detection circuit undetected. Since
the number of these events is quite rare, they will not contribute
much to the failure probability.
It is possible to show that $R\geq t$ and $L\geq t+1$ is necessary
for $t$-strict fault-tolerance for a code that can correct $t$
errors by continuing the same argument\footnote{However, the
standard verification stage would need additional error detections
on the bottom ancilla pair for $R=t$ and $L=t+1$ to be both
necessary and sufficient for $t$-strict fault-tolerance.}. For a
specific $t$-error-correcting code the actual values required for
$R$ and $L$ depend on how each encoding circuit propagates errors.

Let us review why the extended rectangle is the central object in a
fault-tolerance analysis. An encoded circuit where the physical
gates (0-Ga) have been replaced by rectangles can also be viewed as
an encoded circuit with 0-Ga's with a different error model. This
can be achieved simply by inserting perfect decoder-encoder pairs
between the rectangles, see \cite{AGP:ft}. In an ex-Rec with
malignant faults, the rectangle will correspond to a faulty 0-Ga,
whereas for benign faults the rectangle will correspond to a perfect
0-Ga. The reason that one has to take into account an ex-Rec and not
merely a Rec is that faults in the leading 1-EC can combine with
faults in the 1-Rec to produce malignant faults. For example, a
single error in a leading 1-EC does not cause a failure of the
rectangle in which it is contained; this error however can combine
with later errors to give rise to a logical error. At the same time
the presence of a logical error in the leading 1-EC which maps one
codeword onto another will not affect the failure rate of the
rectangle that comes after it since the state which enters this
rectangle is a codeword. Hence in the extended rectangle method
there is no double-counting of errors. Instead, it is an efficient
method to handle the effect of incoming errors and is likely to give
a very tight estimate of the threshold if no other assumptions or
simplifications are present.

In principle one may think that one would also need to be careful
about the effect of incoming errors {\em into} the ex-Rec; perhaps
an incoming error could combine with a seemingly benign fault in the
ex-Rec and give rise to an incorrect rectangle. Thus perhaps one has
to consider the malignancy of sets of faults given a possible {\em
worst case} input to the extended rectangle.

However, one can argue for stabilizer codes and for {\em
deterministic} (to be defined below) error-correction that
malignancy does not depend on incoming errors to the ex-Rec. To show
this, let us first review the formalism of stabilizer codes. A
stabilizer code is the $+1$ eigenspace of an Abelian subgroup of the
Pauli group $P_n$ which contains all $n$-qubit tensor-products of
the Pauli operators $\{X,Y,Z,I\}$. The normalizer $N(S) \subseteq
P_n $ of $S$ is defined as $N(S)=\{E| \forall s \in S, E s
E^{\dagger} \in S \}$. For Pauli operators (which either commute or
anti-commute with each other) $N(S)$ is the simply the group of
Pauli operators that commute with any element in $S$. Any element of
$N(S)\backslash S$ is a logical operator mapping codewords onto each
other. All other Pauli operators $P \notin N(S)$ anti-commute with
at least one element in $S$ and map a code word outside the code
space indicated by a non-zero syndrome. Thus the Pauli group $P_n$
can be partitioned into cosets of $N(S)$ and each of these cosets is
labeled by a different syndrome. The lowest-weight member of each
coset is called the coset leader. Standard syndrome decoding finds,
for each given syndrome, a coset leader with lowest weight and
chooses this as the error correction. Thus the low-weight
(non-degenerate) correctable errors correspond to distinct syndromes
whose coset leader corrects the error. For high-weight errors $E_i$,
all we can say is that $E_i E_{\rm correct} \in N(S)$ since $E_i$
and $E_{\rm correct}$ have the same syndrome.


Now let us consider the issue of incoming errors to an ex-Rec and
assume the following properties of stabilizer error correction.
First, we assume that the part of the 1-EC circuit which couples any
ancillas to the incoming data is {\em deterministic}, i.e. does not
depend on any incoming error on the data. The choice of which
ancillas to couple may depend on some error detection or ancilla
verification. This property holds for many but not all
error-correction circuits; it does not hold, for example, when the
number of repetitions of syndrome extraction depends on the value of
these syndromes. This property {\em does} hold for the circuits used
in this paper. Furthermore, given a stabilizer $S$ and the incoming
error $E_{\rm in}$ on an encoded state, let the 1-EC be such that
the syndrome of the 1-EC uniquely determines in which coset of
$N(S)$ in the Pauli group the error $E_{\rm in}$ lies. In this sense
the 1-EC must be complete error correction for the code that is
used. For example, if for a CSS code the 1-EC only does $Z$ error
correction whereas $X$ errors can map the state outside the code
space, the syndrome information effectively partitions the Pauli
group into cosets of $N(S_X) P_n(X)$. Here $P_n(X)$ is the subgroup
of Pauli operators that only contain $X$ and $I$ and $N(S_X)$ is the
normalizer of the stabilizer subgroup $S_X$ with only $X$ and $I$
Pauli operators. In this case the syndrome does not uniquely assign
the incoming error to a coset of $N(S)$. Thirdly, upon any incoming
error $E_{\rm in}$ a perfect 1-EC determines a syndrome that
corrects $E_{\rm in}$ modulo a logical error (given by an element in
$N(S)$). This is a basic property of stabilizer error correction as
described above.

Let then the incoming state to an ex-Rec be a state in the
code-space of the stabilizer with an additional error $E_{\rm in}$.
We want to show that the state that comes out of the leading 1-EC is
again some state in the code space with an additional error $E_{\rm
out}$ that only depends on the errors inside the 1-EC, $E_{\rm ec}$,
i.e $E_{\rm out}=f(E_{\rm ec})$ where $f$ is independent of $E_{\rm
in}$. Any 1-EC circuit for stabilizer codes can be implemented with
Clifford gates. Given an incoming error $E_{\rm in}$ and error
inside the 1-EC $E_{\rm ec}$, it follows (because a 1-EC for any
stabilizer code can be implemented with Clifford gates) that the
1-EC has syndrome $s(E_{\rm in} h_1(E_{\rm ec}))$ where $h_1$ is a
function independent of $E_{\rm in}$. Based on the syndrome the
correction step will be some $E_{\rm correct}= E_{\rm in} h_1(E_{\rm
ec}) \mod N(S)$. Before error correction the data has error
$h_2(E_{\rm ec}) E_{\rm in}$ where $h_2(E_{\rm ec})$ is the part of
$E_{\rm ec}$ that has propagated to the data. After error correction
the data thus has error $h_2(E_{\rm ec}) h_1(E_{\rm ec}) \mod N(S)$.
We strip off the logical error in $N(S)$ and identify $E_{\rm
out}=h_2(E_{\rm ec}) h_1(E_{\rm ec})$. Note that when the EC is not
deterministic, the functions $h_1$ and $h_2$ can depend on $E_{\rm
in}$. \qed


We discuss the explicit decoding of the error syndromes for each
code in Appendix \ref{app:decode}.

\subsection{Monte-Carlo Implementation of Method}
\label{sec:mcmethod}


Given the AGP method the numerical problem to be solved is whether a
Rec is correct given a set of faults in the ex-Rec containing it.
This set of faults is generated using depolarizing noise with error
probability $p_0$ for each location in the circuit. We calculate the
failure rate of the ex-Rec, i.e. the probability that the Rec is not
correct, for fixed $R$ and $L$. This implies that sometimes there
are no verified ancillas available for a 1-EC. If this happens for
{\em any} of the 1-ECs inside the extended rectangle, we call this a
failure of the extended rectangle. We do this for all codes except
for Reichardt's use of [[49,1,9]]. The reason for this exception is
that for Reichardt's method the failure rate of ancillas may be
rather high. If we let failure of having verified ancillas in the
leading 1-EC determine failure of the rectangle after this 1-EC, we
are possibly double-counting errors. Hence in Reichardt's method we
replace any failed ancillas in the leading 1-EC by a perfect ancilla
and do not call failure. As the results show, even under this
assumption, the [[49,1,9]] concatenated code with error-detection
and finite resources is not a great performer.

In general, our assumption on the effect of failed ancilla
preparations may make our estimates for the pseudo-threshold for the
EC circuits slightly more pessimistic.

We will estimate the failure rate of a CNOT ex-Rec, since this is by
far the biggest circuit among the Clifford ex-Recs. As we argued in
Section \ref{sec:univ}, the non-Clifford (and possibly other
Clifford) gates will be implemented via injection-and-distillation
so that their implementation will not affect the threshold. Pauli
gates are not applied within a Clifford ex-Rec because they can be
stored in classical memory as the Pauli frame and applied only prior
to the execution of non-Clifford gates.

Given a fixed $R$ and $L$, we will estimate the failure rate
$p_1(p_0)=\frac{N_{\rm fail}}{N}$ where $N_{\rm fail}$ is the
number of Monte-Carlo samples that fail (i.e. the number of times
we simulate the extended rectangle with randomly generated faults
and observe that the rectangle is incorrect) and $N$ is the total
number of runs. With high probability this estimated $p_1$ lies
within one standard deviation of the real $p_1$. In this way we
collect data points $p_1(p_0)$ for different values of $p_0$. We
then take these points as the mean of a normal distribution for
each $p_0$. We sample from these normal distributions and for each
set of samples we determine a small degree polynomial $p_1(p_0)$
fitting the samples.
The equation $p_1(p_0)=p_0$ gives us a sample of the threshold and
we put an error bar on this result by calculating the standard
deviation of the obtained threshold samples.


The way we test for correctness of a rectangle for a given pattern
of faults in the ex-Rec is as follows: Let $E_{\rm out}$ be the
outgoing error of the leading 1-EC. We use syndrome decoding to
determine the coset leader $E_{\rm lead}$ corresponding to the
coset of $N(S)$ in the Pauli group of this $E_{\rm out}$. We
propagate this $E_{\rm lead}$ through the rectangle, let $f(E_{\rm
lead})$ be the outgoing error on the data. We follow the rectangle
by an ideal decoder and let $E_{\rm correct}$ be the correction
suggested by the ideal decoder. Then we test whether $E_{\rm
correct} f(E_{\rm lead})$ commutes with both $\overline{X}$ and
$\overline{Z}$. If it does, we infer that no logical faults
occurred, hence the rectangle was correct. Otherwise we call
failure.


An alternative way of using the AGP correctness criterion is to
count or sample malignant fault sets. This method is advantageous
if one wants to estimate the threshold for worst-case adversarial
noise. In such an application, one fixes the number of faults and
counts or samples how many sets with this fixed number of faults
are malignant. The failure rate $p_1$ is a polynomial in $p_0$
with factors that are determined using the malignant set counts
(or estimates of these counts determined by sampling). For codes
with large distance this method becomes cumbersome, as the total
number of possible fault-sets of size $t+1$, ${A \choose t+1}$
becomes large. Here $A$ is the total number of locations in the
ex-Rec. This sampling method is difficult but still possible for
the Golay code, but it is not possible for codes of higher
distance.

The advantage of the malignant set counting or sampling method is
that one gets an upper bound on $p_1$ for arbitrary values of $p_0$.
This makes it possible to estimate $p_1$ even for very small values
of $p_0$. We use the Monte-Carlo simulation to estimate $p_1$, but
the number of samples required becomes quite large if one wants to
estimate $p_1$ with good relative error for small $p_0$. In such
cases we extrapolate the values for $p_1$ obtained from larger
values of $p_0$, see Section \ref{sec:results_pseudo}.


\subsection{Software and Computer Use}
\label{sec:software}

On the website \cite{cross:tools} one can find a set of software
tools that have been developed for this and other future
fault-tolerance projects. The quantum circuits for the CNOT ex-Recs
based on CSS codes are highly structured and can be mechanically
assembled in $O(n^3)$ time for block-size $n$ given the classical
codes $C_1$ and $C_2$. We have used MAGMA \cite{MAGMA} and/or GAP
\cite{GAP} (using the GUAVA package \cite{GUAVA}) to construct
quantum codes and compute their parameters. The code stabilizers are
copied from the computer algebra programs into our circuit synthesis
and simulation programs, where they are again verified to have the
required commutation relations.

The simulation and circuit synthesis programs are implemented in C++
and use MPI \cite{MPI} for communication during embarrassingly
parallel tasks. The project is entirely open source and makes use of
preexisting open source libraries such as a Galois field
implementation \cite{Arash} and a weighted matching algorithm
\cite{thesis:rothberg}. Importantly, the same functions and
procedures are used in the Monte-Carlo simulation. This gives us
increased confidence in the simulation output.

The symmetries of the pair count matrix for some distance-3 code
circuits
and the lack of single-location malignancies in all circuits
strongly suggests that our circuit constructions are indeed
fault-tolerant against single errors. Furthermore, we strictly check
all input and intermediate results for consistency at runtime. The
programs can be optimized and further improved, but we leave this to
future work and encourage development by making the code publicly
available \cite{cross:tools}.

The simulations were carried out on a relatively small allocation
of Blue Gene L at the IBM T. J. Watson Research Center. Typically
we used between 64 and 256 PowerPC 440 700 MHz CPUs. Each pair of
CPUs had access to 512 Mb of local memory. Using 256 CPUs gave us
roughly a factor of 50 speed-up over a typical single-processor
desktop machine. The entire process of development and debugging
took many months, but we estimate that all of the data could be
retaken in several weeks with these computing resources.

\begin{figure}[htb]
\begin{center}
\input{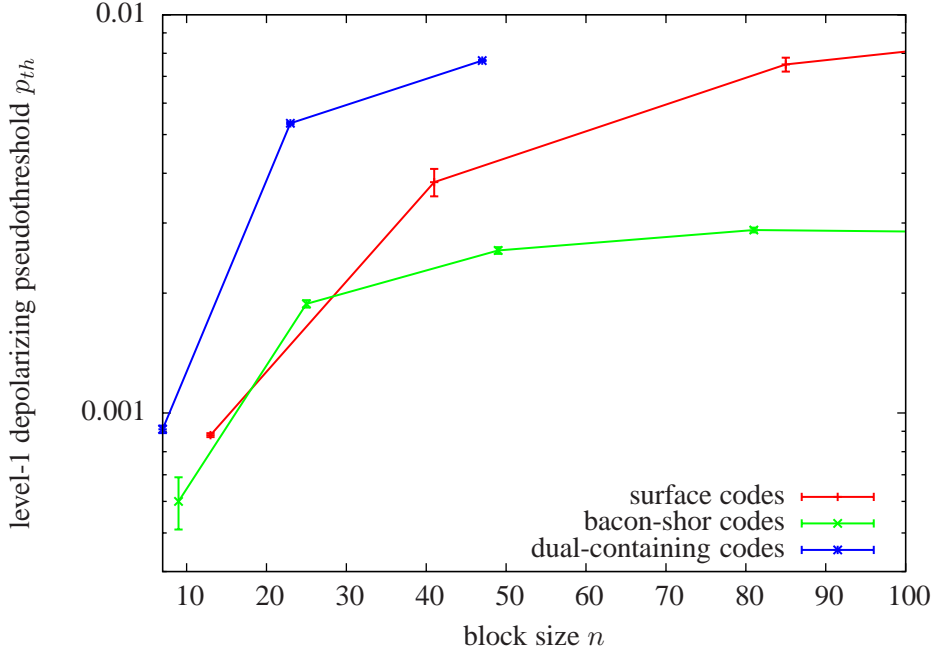}
\end{center}
\caption{Level-1 depolarizing pseudo-threshold for three families
of codes with perfect ancillas for Steane-EC: surface codes,
dual-containing codes, and Bacon-Shor codes.  This plot indicates
that under no circumstances can thresholds reach $1\%$ for the
codes in our study. The data points are connected by lines merely
as a guide to the eye. } \label{fig:perfect}
\end{figure}

\section{Results}
\label{sec:results}

Tables~\ref{table:complete1}, \ref{table:complete2},
\ref{table:complete3}, \ref{table:surfacedat}, and
\ref{table:perfectdat} list the complete set of results of our
studies. Our results are obtained assuming that all locations
including memory locations suffer from noise at the same noise rate,
unless specified otherwise.

\subsection{Perfect Ancillas}

\begin{figure}[htb]
\begin{center}
\input{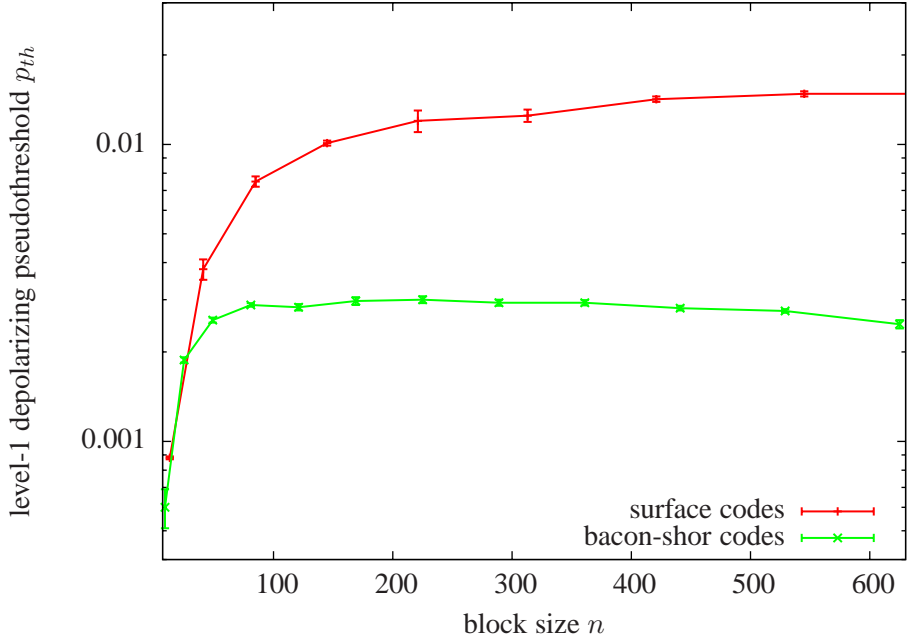}
\end{center}
\caption{Level-1 depolarizing pseudo-threshold for surface codes and
Bacon-Shor codes using perfect ancillas for Steane-EC.}
\label{fig:perfectlonger}
\end{figure}

In our first study, and only in this Section, we assume that
ancillas for Steane error correction can be prepared flawlessly, see
Figure \ref{fig:perfect}. In such a scenario, the threshold is
largely determined by the error-correction properties of the code
(see also the analysis in \cite{eastin}), in particular its
(effective) distance. For families of quantum error-correcting codes
in which the effective distance is linear in the block-size, we
expect the threshold to be a monotonically increasing as a function
of $n$, see Section \ref{sec:scaling}. In Figure~\ref{fig:perfect}
and Figure \ref{fig:perfectlonger} we have plotted the
pseudo-thresholds for three families of codes: surface codes, some
dual-containing codes, and Bacon-Shor codes. The surface codes and
Bacon-Shor codes apparently have fairly good distance properties,
even though there is some decline in the BS code family for large
$n$.
Figure~\ref{fig:perfect} shows we cannot expect a threshold over
$1\%$ for the codes we have studied using Steane-EC -- introducing
noise realistically into the ancilla preparation circuits cannot
increase the pseudo-threshold. Note that if we do Shor-EC on the
surface codes we cannot expect thresholds exceeding about $3\%$, see
the arguments in \cite{dennis+:top}.

When we assume that the logical Bell pairs of Knill's circuit can be
prepared flawlessly, the level-1 pseudo-threshold of the
$[[5,1,3]]$'s ${\rm T_3}$ gate is $(2.0\pm 0.1)\times 10^{-4}$
\footnote{The pseudo-threshold in this case is the point at which
the failure rate $p_1$ of a ${\rm T_3}$ ex-Rec is the same as the
base error rate $p_0$ of all elementary gates in the ex-Rec.}. This
is roughly an order of magnitude below the Steane code with perfect
ancillas.

\subsection{Pseudo-Thresholds}
\label{sec:results_pseudo}

\begin{figure}[ht]
\begin{center}
\input{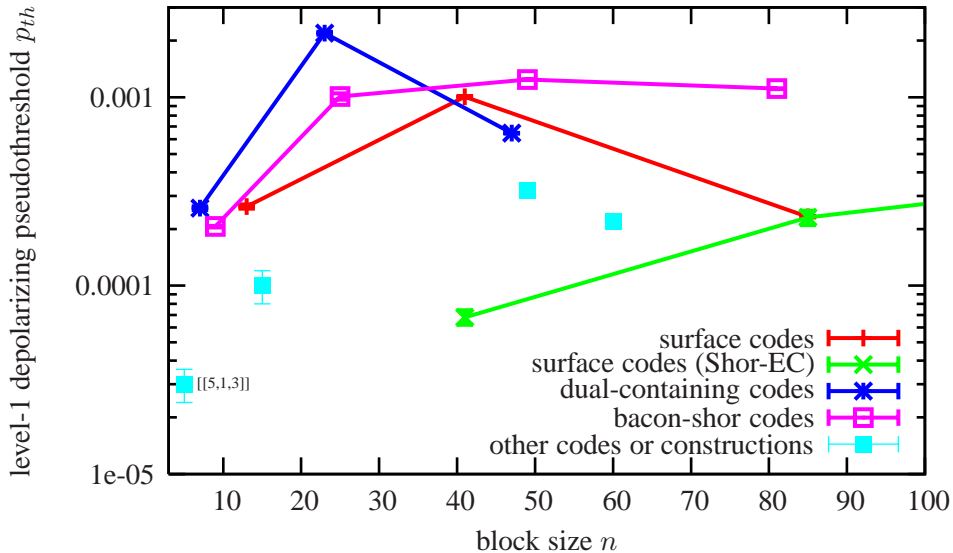}
\end{center}
\caption{Level-1 depolarizing pseudo-threshold versus block size.
The other codes are the $[[5,1,3]]$ non-CSS code, the $[[15,1,3]]$
Reed-Muller code, the $[[49,1,9]]$ (dual-containing) concatenated
Steane code using $L=15$ attempts to prepare using error detection
at level-1, and the $[[60,4,10]]$ (dual-containing) concatenated
polynomial code using $L=20$ attempts to prepare ancillas.}
\label{fig:block}
\end{figure}

\begin{figure}[ht]
\begin{center}
\input{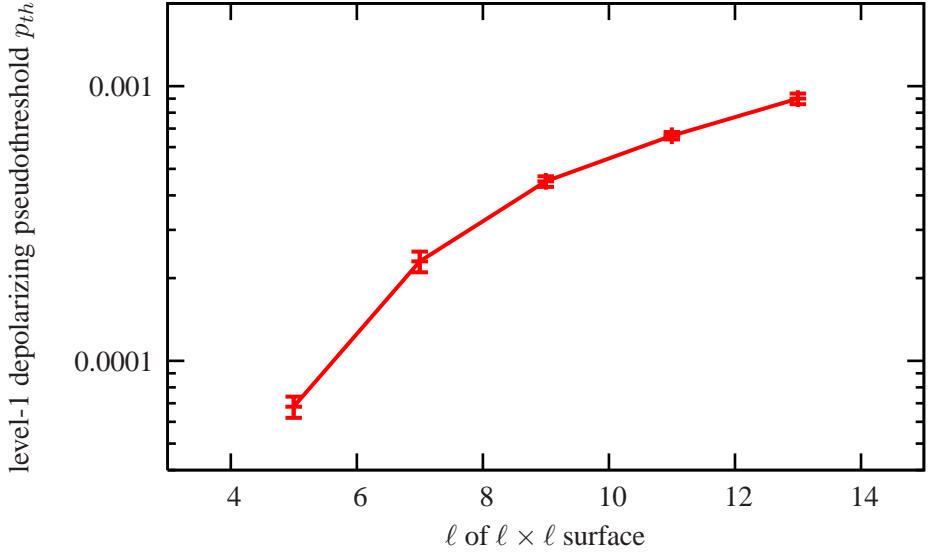}
\end{center}
\caption{Surface code level-1 depolarizing pseudo-threshold versus
$\ell$ for $\ell\times\ell$ surface code (the block-size
$n=\ell^2+(\ell-1)^2$). The ex-Rec is a transversal CNOT gate with
$\ell$ sequential Shor-EC steps per EC. The pseudo-threshold
increases with $\ell$ and is expected to approach a constant value
in the limit of large $\ell$, unlike the other codes in this study.}
\label{fig:surfacethreshold}
\end{figure}

In Figure \ref{fig:block} we tabulate for each code the maximum
pseudo-threshold over the various choices of $R$ and $L$. The
maximum overall pseudo-threshold $(2.25\pm 0.03)\times 10^{-3}$ is
attained by the Golay code with $L=30$ and $R=1$. The two code
families, Bacon-Shor and surface, both attain a peak threshold and
then decline when we use Steane-EC. The peak Bacon-Shor code is the
$[[49,1,7]]$ at $(1.224\pm 0.005)\times 10^{-3}$ with $L=9$ and
$R=1$. The peak surface code (using Steane error correction) is
$[[41,1,5]]$ at $(1.008\pm 0.008)\times 10^{-3}$ at $L=30$ and
$R=1$. Interestingly when we use Shor-EC for the surface codes the
performance is quite different. Shor-EC does not do as well as
Steane-EC for small block sizes, but for larger block size Shor-EC
gives a threshold that asymptotes to a finite value, see Figure
\ref{fig:surfacethreshold}. For small block size the thresholds of
the surface codes are not as good as of some other codes such as the
Golay code and the Bacon-Shor codes.

It is clear from the data that the pseudo-threshold increases with
increasing $L$. Our main interest in this study is in circuits
with small overhead and hence with a relatively small number of
preparation attempts $L$. In various cases the thresholds stated
for finite $L$ will be thus be lower than the one in the $L
\rightarrow \infty$ limit. Notably, this occurs for the
$[[49,1,9]]$ code, where we expect thresholds approaching $1
\times 10^{-2}$ with many more ancilla preparation attempts
\cite{reichardt:concat7}. In other cases one can take the perfect
ancilla results in Figure \ref{fig:perfect} and the Tables as
upper bounds on the $L \rightarrow \infty$ pseudo-threshold.


The use of weakly fault-tolerant circuits, i.e. small $R$ and small
$L$ is meant to get improved threshold behavior for finite
resources. The best performance of a code can be expected for $L
\rightarrow \infty$, since one would always use ancillas which
passed verification. When $L \rightarrow \infty$, it is clear that,
--at least below threshold--, optimal performance is likely to be
achieved when $R$ is taken as large as possible, since then an
ancilla is maximally verified. However, at finite $L$, a larger $R$
will let more ancillas fail and hence increase the chance for an
extended rectangle to fail (remember that we, pessimistically, call
an extended rectangle failed if we don't have ancillas for EC).
Hence for small $L$, small $R$, weakly fault-tolerant verification
circuits can outperform circuits with the same $L$ and larger $R$.

\subsection{Influence of Storage Errors}

\begin{figure}[htb]
\begin{center}
\input{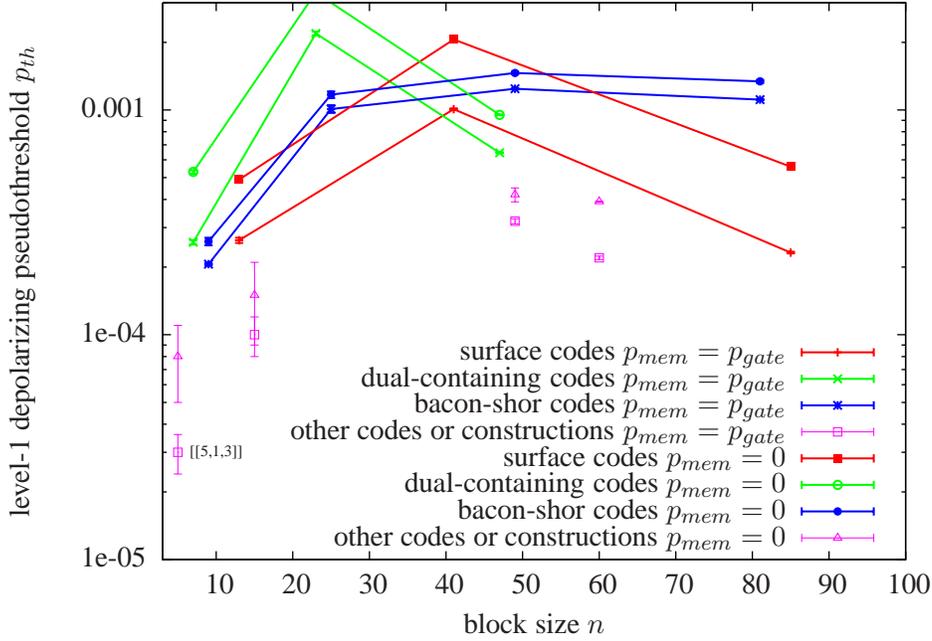}
\end{center}
\caption{Pseudo-thresholds versus block size for Steane-EC and
Knill-EC circuits, comparing the case where the memory failure
rate equals the gate failure rate with the case where the memory
failure rate is zero. Naturally the difference is smallest where
we have taken advantage of simple encoders as those for the
Bacon-Shor codes.} \label{fig:blockcompare}
\end{figure}

In Figure~\ref{fig:blockcompare} we replot the pseudo-threshold
versus block-size when storage error rates (on memory locations)
are zero. The peak pseudo-threshold increases to $(3.33\pm
0.02)\times 10^{-3}$. The Figure shows that storage errors do not
influence the pseudo-threshold appreciably. The Bacon-Shor codes
are least affected by storage errors because the encoding circuits
are extremely simple. The non-CSS $[[5,1,3]]$ code is most greatly
affected because storage errors can enter into the ${\rm T_3}$
gate sub-circuit, the $|\overline{0}\rangle$ encoders, and the
cat-state encoders at many locations.

\subsection{Logical Error Rate versus Overhead}

\begin{figure}[ht]
\begin{center}
\input{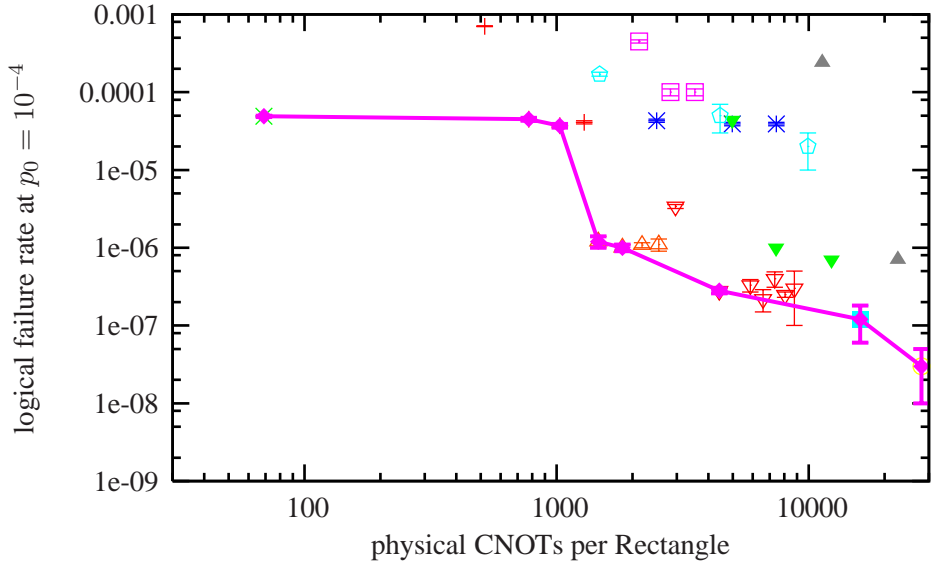}
\end{center}
\caption{Level-1 logical error rate (probability of failure of a
CNOT ex-Rec) versus the number of CNOTs per rectangle. The line
connects the points of the best performing codes. Points with the
same shape (color online) belong to the same code but have
different circuit parameters. The error rates are evaluated at a
fixed $p_0=10^{-4}$.  The results C4/C6 scheme of
\protect{\cite{knill:nature}} are shown for comparison.}
\label{fig:errorrate_full}
\end{figure}

\begin{figure}[htb]
\begin{center}
\input{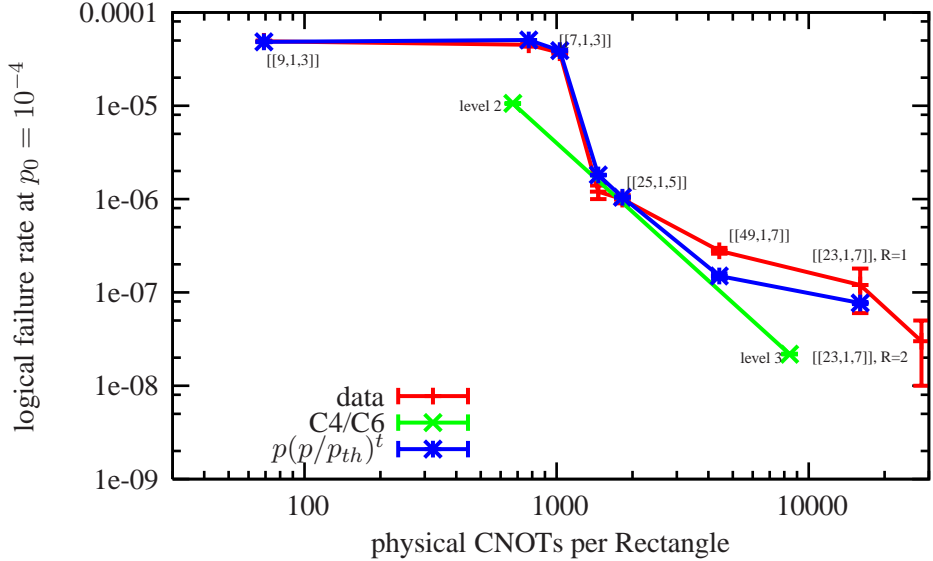}
\end{center}
\caption{Level-1 logical error rate (probability of failure of CNOT
ex-Rec) versus the number of CNOTs per rectangle for the best
performing codes. The subset of data plotted here was chosen so that
the error rate decreases monotonically with the rectangle size and
there is no code with lower error rate at a given rectangle size.
The error rates are evaluated at a fixed $p_0=10^{-4}$.}
\label{fig:errorrate}
\end{figure}

\begin{figure}[htb]
\begin{center}
\input{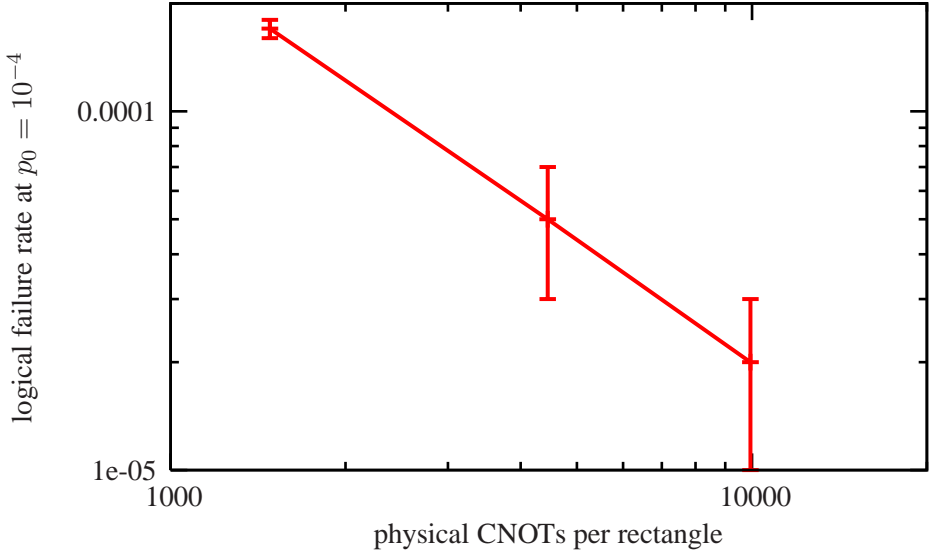}
\end{center}
\caption{Level-1 logical error rate versus the number of CNOTs per
rectangle for the $\ell\times\ell$ surface codes, $\ell=5,7,9$. It
is expected that the error rate decreases exponentially as $\ell$
increases for fixed $p_0=10^{-4}$. } \label{fig:surfaceerrorrate}
\end{figure}

The threshold is an extremely important figure of merit for
fault-tolerant circuit constructions. But practically speaking, we
are also interested in how quickly the error rate decreases if the
initial error rates are low enough for a given overhead. Figures
\ref{fig:errorrate_full}, \ref{fig:errorrate}, and
\ref{fig:surfaceerrorrate} plot the probability of failure of a CNOT
ex-Rec (defined in Section \ref{sec:mcmethod}) versus the number of
physical CNOTs in a rectangle at $p_0=10^{-4}$.
Even though there are other measures of overhead, such as total
number of qubits involved in the rectangle or the depth of the
rectangle circuit, we have chosen the number of CNOTs per rectangle
as an estimate for the overhead since it approximately captures the
total size, i.e. depth times width, of the rectangle.

The Golay code achieves the lowest logical error rate for codes with
fewer than $O(10^{4})$ CNOT gates per rectangle, and that rate can
be further reduced by increasing the number of verification rounds
to $R=2$. There is a clear tradeoff between the number of physical
CNOTs per rectangle and the logical error rate.
We note that given the lack of code specific optimizations, the
achievable overheads for various codes may be somewhat less than
what is estimated here. For the Golay code and the Bacon-Shor codes
for example, the overhead may come down by at least a factor of 2 by
using simplified verification circuits. Viewed an a log-scale such
decrease in resources is relatively small. We also see in Figure
\ref{fig:errorrate} that the approximate expression for the failure
rate, Eq. (\ref{eq:p1estim}), gives a pretty good estimate of the
actual failure rate.

Some of the error rates plotted in Figure~\ref{fig:errorrate_full}
were extrapolated from error rates at higher values of $p_0$ and may
only be rough indications of the actual error rates. For
small values of $p_0$ the logarithm of the error rate $p_1(p_0)$ is
expected to be approximately linear in $p_0$. We extrapolate from a
least-squares fit to this line.
Tables~\ref{table:complete1}, \ref{table:complete2}, and
\ref{table:complete3} indicate these extrapolated rates by enclosing
them in square brackets. The extrapolations are only plotted for the
$5\times 5$ surface code and the $9\times 9$ Bacon-Shor code and are plotted
without errorbars for these two points
\footnote{Unfortunately, using the same method by
which we obtain error estimates for our calculated pseudothresholds, we find error
estimates that are an order of magnitude larger than these extrapolated values.
In the few places where we make these extrapolations, the results should only be
taken as rough indications of the actual error rates the codes can attain.}.

For the Golay code we have looked at the behavior of the threshold
for $R=1, 2, 3$. One important empirical observation is the
following. The pseudo-threshold can increase slightly while the
logical error rate for $p_0=10^{-4}$ remains the same. This
happens for the Golay code when $R=1$ and $L$ is increased from
$10$ to $20$. Furthermore, the pseudo-threshold can decrease while
the logical error rate decreases too. This also happens for the
Golay code when $L=10$ and $R$ is increased from $1$ to $2$.
This suggests that the pseudo-threshold value is sensitive to
higher order effects that quickly become negligible at lower error
rates. Thus a desired logical error rate may be achievable with
significantly fewer ancilla resources $L$ than are necessary to
maximize the pseudo-threshold, provided the initial error rate
$p_0$ is not too close to the pseudo-threshold.

In Figure \ref{fig:errorrate} we have also added Knill's $C_4/C_6$
Fibonacci scheme \cite{knill:nature} at 2 and 3 levels of encoding.
These data points are derived from his paper \footnote{Note that his
error model is slightly different from ours but we take the
dominating physical CNOT error rate to be the same.}. At level 2 the
detected error rate of the logical CNOT is $(1.06\pm 0.01)\times
10^{-5}$ and at level 3 the detected error-rate is $(2.18\pm 0.02
)\times 10^{-8}$.

The plot shows that [[9,1,3]] is still better than the $C_4/C_6$
scheme in terms of overhead, but the $C_4/C_6$ Fibonacci scheme
definitely beats [[7,1,3]]. The next two Bacon-Shor codes fill a
void between $C_4/C_6$ level 2 and $C_4/C_6$ level 3.

For the surface codes (see Fig. \ref{fig:surfaceerrorrate}) we
note that the error rates are relatively high compared to other
error-correcting codes with comparable numbers of CNOTs per
rectangle. However one should remember that the circuits for the
surface codes are already spatially local in two dimensions
whereas the circuits for any of the other codes, for example, the
Golay code, are not.

\section{Discussion}

In our study we have considered bottom codes and their performance
in a bottom-top code architecture. Our best threshold around
$2\times10^{-3}$ is seen for the Golay code, and many other codes
both larger and smaller were studied and found to have much worse
thresholds. An important figure of merit is the logical error rate
versus overhead curve which shows that the Bacon-Shor codes are
competitive with Knill's $C_4/C_6$ scheme at a base error rate of
$10^{-4}$.

We have seen that the constraint of finite resources, i.e. limited
$R$ and $L$ in Steane EC, can considerably and negatively impact
noise thresholds. An example is Reichardt's estimate for
$[[49,1,9]]$ when $L \rightarrow \infty$ versus our estimates for
this construction at small $L$. For code families with low-weight
stabilizers, Shor EC may give rise to thresholds which grow with
block-size. For code families which do not have this property, e.g.
general quantum BCH codes, the limit resource constraint on $R$ and
$L$ and the complexity of the encoding circuits start pushing the
thresholds down beyond some peak performance block-size.

In this landscape of codes and their performances, one of the
missing players is the surface code scheme of \cite{RHG:topo} in
which many qubits are encoded in one surface code and the CNOT
gate is done in a topological manner. In principle, the possible
advantage of this scheme is that if one uses enough space (meaning
block size) one would reach the asymptotic threshold of a simple
EC rectangle (no 1-Ga). We have in fact analyzed an ex-Rec where
the Rec is only Shor-EC on a $\ell \times \ell$ surface and we
find that this asymptotic memory threshold for $\ell \rightarrow
\infty$ is about $3.5 \times 10^{-3}$. This is a factor of two
lower than the number stated in \cite{RHG:topo}. For finite block
size one could analyze a CNOT ex-Rec for this topological scheme
just as for the other codes. Like all the other codes, the
topological scheme will have a trade-off between overhead and
logical error rate. It will be interesting to see whether topology
and block coding provide an efficient way of using resources and
how it compares to a local version of a bottom-top architecture
discussed in this paper.


For a bottom-top architecture it will be important to study the
performance of top codes in order to understand at what error rate
one should switch from bottom to top code and what total overhead
one can expect. Concerning a choice of top code we expect the
following. First of all, given the constructions of
\cite{SI:networks}, one can expect that a $[[n,k,d]]$ block code
has a threshold comparable to a $[[n,1,d]]$ code. Secondly, the
networks in \cite{SI:networks} show how to do logical gates on
qubits inside the block codes using essentially gate-teleportation
and Knill-EC. One issue of concern for block codes is the
complexity of the encoding circuit as a function of block size. It
would be highly desirable to consider block codes with EC circuits
that are linear in $n$, otherwise one would expect the threshold
to decline as a function of $n$.

There is another desirable property of top codes which relates to
the transversality of gates. In order to minimize overhead, it is
desirable that the T gate is transversal for the top code. The
reason is as follows. In order to have maximal freedom in picking a
bottom code we will only require that it has a transversal ${\rm
CNOT}$. Thus all other gates, in particular ${\rm T}=e^{i \pi Z/8}$
and the phase gate S, should be either performed by more complicated
fault-tolerant 1-Ga or be implemented by the
injection-and-distillation scheme. If the fault-tolerant circuits
for these non-transversal gates have poorer thresholds than the CNOT
gate, then the injection-and-distillation scheme is the preferred
solution. In the injection-and-distillation scheme, the obtained
error rates of the encoded and distilled ancillas will be limited by
the noise rates on the Clifford gates which distill the ancillas,
since the Clifford distillation circuit is not fault-tolerant.
Assume we teleport the ancillas into $C_{\rm top} \circ C_{\rm bot}$
and get Clifford gates with $O(10^{-15})$ error rate. Since a
circuit such as Bravyi-Kitaev distillation uses $O(10^{3})$ gates,
the error rates of the distilled ancillas can be as high as
$O(10^{-11})$. Thus by these schemes the, say, T error rate is
always trailing the transversal gate error rates. But assume that
the T gate is transversal for the top code and thus we only inject
the T ancillas into $C_{\rm bot}$. Then even though the once encoded
gate $C_{\rm bot}({\rm T})$ has an error rate of, say, $O(10^{-4})$,
the twice-encoded gate $C_{\rm top} \circ C_{\rm bot}({\rm T})$ will
mostly likely have an error rate similar to other Clifford gates
since there are very few $C_{\rm bot}({\rm T})$ in the twice-encoded
gate compared to the EC parts.
Of course, the top code will have other non-transversal gates; for
example the [[15,1,3]] code has a transversal T gate but not a
transversal Hadamard gate. If the bottom code has a transversal
Hadamard gate, we can implement a fault-tolerant H in $C_{\rm top}
\circ C_{\rm bot}$ by using the fault-tolerant non-transversal
gadget for the Hadamard gate in $C_{\rm top}$ and implementing the
resulting Clifford gates. This shows that there are possible
constructions which would allow all gates needed for universality to
be implemented with approximately the same, low, error rate, while
the noise threshold of such scheme is determined by the noise
threshold for the transversal Clifford gates.


\section{Acknowledgements}
We acknowledge and thank Panos Aliferis, Sergey Bravyi and Austin
Fowler for various fruitful discussions and insights. AC thanks
Isaac Chuang for helpful discussions. DPD and BMT acknowledge
support by DTO through ARO contract number W911NF-04-C-0098. AC is
grateful for partial support from the Urabe Crest Project of the
Japan Science and Technology Agency.

\appendix


\section{Various Aspects of Steane-EC}
\label{app:css}

A binary $[[n,k,d]]$ CSS code $\textrm{CSS}(C_1,C_2)$ is constructed
using two classical linear error correcting codes $C_2^\perp\subseteq C_1$
and has the codewords:
\begin{equation}
\ket{\overline{a}} = \frac{1}{\sqrt{|C_2^\perp|}} \sum_{c\in C_2^\perp}
\ket{c+a}\ \textrm{where}\ a\in C_1/C_2^\perp.
\end{equation}
Each row $r$ of the parity check matrix of $C_2$ gives the
stabilizer generators $X(r)$, and each row $s$ of the parity check
matrix of $C_1$ gives stabilizer generators $Z(s)$, where
$U(r)=U_1^{r_1}\otimes\dots\otimes U_n^{r_n}$. It is easy to check
that $\ket{\overline{a}}$ is a simultaneous eigenstate of these
stabilizer generators: (1) a row $r$ of the parity check matrix of
$C_2$ must be an element of $C_2^\perp$, so adding it to each
codeword in the superposition $\ket{\overline{a}}$ leaves the state
unchanged, and (2) every codeword in the superposition
$\ket{\overline{a}}$ is an element of $C_1$, so it must pass the
parity checks of $C_1$. A basis of the $2^k$ cosets of $C_2^\perp$
in $C_1$ corresponds to logical $X$ operations $\overline{X}(a)$ on
the code space because
$\overline{X}(a)\ket{\overline{0}}=\ket{\overline{a}}$. Similarly, a
basis of the $2^k$ cosets of $C_1^\perp$ in $C_2$ corresponds to
logical $Z$ operations $\overline{Z}(b)$ since
$Z(b)\ket{\overline{a}}=(-1)^{b\cdot a}\ket{\overline{a}}$. We can
choose these bases such that the logical operators obey the
commutation relations of the $k$-qubit Pauli group.

A special case of the CSS construction occurs when
$C_2^\perp=C_1^\perp$, in which case $C_1$ is a dual-containing
code. The X and Z stabilizer generators have identical supports and
the Hadamard H gate is transversal.
If in addition the weight of each stabilizer generator is a multiple of
$4$, $C_1^{\perp}$ is called doubly-even and the quantum code has a
transversal S gate.
The code does not have any transversal gates outside of the Clifford group \cite{rains:d2}
in this case.




Steane error correction for a CSS code $CSS(C_1,C_2)$,
$C_2^\perp\subseteq C_1$ uses $|\overline{+}\rangle$ and
$|\overline{0}\rangle$ ancilla states. These states can be encoded
directly from the generator matrices of $C_1$ and $C_2^\perp$,
respectively, according to a well-known procedure. The generator
matrix $G$ has $n$ columns and $k_1$ rows for $C_1$ or
$n-k_2$ rows for $C_2^\perp$, and the quantum code encodes $k=k_1+k_2-n$
qubits. Gaussian elimination puts a generator matrix into
standard form $G=(I |A)$ where $I$ is an identity matrix and $A$ is
a binary matrix. The $i$th row of the generator matrix specifies the
controls and targets of $w_i$ CNOT gates, where $w_i$ is the weight
of the row minus one. In the next section we discuss how to
implement this circuit in a way that minimizes the number of memory
locations. The depth of the resulting CNOT circuit is
$w=\textrm{max}\{w_i\}$, assuming equal cost for any pair of qubits
to communicate. The number of fault locations in an encoder is
summarized by the following expressions:
\begin{align*}
A_\textrm{enc}(n,k_1,k_2,w) & \leq n + w\textrm{max}(k_1,n-k_2),\ \textrm{no memory noise} \\
A_\textrm{enc}(n,k_1,k_2,w) & \leq n + w n,\ \textrm{memory noise}.
\end{align*}
For particular states, different scaling is possible. For example,
for the Bacon-Shor codes one can make the encoded ancillas using
$O(n)$ 2-qubit gates.
In general, any unitary stabilizer circuit has an equivalent circuit
with $O(n^2)$ gates and $O(\log{n})$ depth \cite{AG:stabilizer}.



One method of verifying the encoded ancilla against low-weight
correlated errors is to use transversal gates to perform error
detection. One possible error detection method consumes three
additional ancilla and uses 3 transversal CNOT gates and 3
transversal measurements.
The cost of verifying is:
\begin{align*}
A_\textrm{ver}(n,k_1,k_2,w,R) & \leq R(3A_\textrm{enc}(n,k_1,k_2,w) + 6n),\ \textrm{no memory noise} \\
A_\textrm{ver}(n,k_1,k_2,w,R) & \leq R(3A_\textrm{enc}(n,k_1,k_2,w) + 6n) + n,\ \textrm{memory noise}.
\end{align*}
Again, these expressions assume equal cost for any pair of qubits to communicate.

Finally, we can write expressions for the total number of fault
locations in a CNOT extended rectangle using Steane error
correction:
\begin{equation*}
A(n,k,w,R) \leq 8A_\textrm{enc}(n,k_1,k_2,w) + 8A_\textrm{ver}(n,k_1,k_2,w,R) + 17n.
\end{equation*}
If we set $R=t$ then the total number of fault locations is
$A(n,k_1,k_2,w,R)=O(wnt)$ using this method of error correction and
assuming equal communication costs between qubits. In the worst case
this can be $O(n^3)$.

\subsection{Latin Rectangle Method for Optimizing Encoding Circuits}
\label{app:latin}

There is a simple method for minimizing the number of memory
locations in an ancilla encoding circuit due to Steane. Steane
puts the generator matrix $G$ of a linear binary code into
standard form $(I|A)$ using Gaussian elimination. An encoding
circuit for the logical zero state can be constructing by looking
at the $A$ matrix for the code $C_2^\perp$. Every $1$ in the $A$
matrix gives a CNOT gate in the encoder. The control qubits are
the $1$s in the $I$ part of $G$ and the target qubits are the $1$s
in the $A$ part of $G$.

For example, we have $G = (1010101,0110011,0001111)$ for the
$[7,3,4]$ code, which is the $C_2^\perp$ for Steane's [[7,1,3]]
code. Transposing columns 3 and 4 gives the standard form and an $A
= (1101,1011,0111)$. This means there are 9 CNOT gates in the
logical zero encoder. We can assign each CNOT a time-step so that no
qubit is involved in two gates at once. That constraint makes a
time-step assignment the same as finding a partial Latin rectangle.
The Latin rectangle to complete is
$\left[\begin{array}{cccc} ? & ? & & ? \\
? & & ? & ? \\ & ? & ? & ? \end{array}\right]$ and one possible completion
is
$\left[\begin{array}{cccc}
1 & 2 &   & 3 \\
2 &   & 3 & 1 \\
  & 3 & 1 & 2
\end{array}\right]$.
The circuit corresponding to the time-step assignment is:
\begin{verbatim}
 # time 1
 cnot 1,4
 cnot 2,7
 cnot 3,6
 # time 2
 cnot 1,5
 cnot 2,4
 cnot 3,7
 # time 3
 cnot 1,7
 cnot 2,6
 cnot 3,5
\end{verbatim}
We have to undo the qubit permutation that occurred in the Gauss
elimination to standard form, so at the end we should switch back qubits 3 and 4.
This is the smallest depth $(3)$ that a circuit for $A$ can have.
The smallest depth is the maximum row or column sum $w$ of $A$.

The problem of completing the Latin rectangle and therefore of
computing the optimal time-step assignment for a matrix $A$ is
equivalent to edge coloring a bipartite graph with the minimum
number of colors. We construct the graph in the following way. The
left set of vertices corresponds to the control qubits. The right
set of vertices corresponds to the target qubits. A control and
target vertex are connected by an edge if there is a CNOT between
those two qubits. Assign a color to an edge to indicate what
time-step we plan to do that CNOT gate. Since we cannot have two
CNOT gates occur at the same time using the same qubit, all of the
edges incident to a given vertex must have different colors. This
means that a valid schedule corresponds to an edge coloring of
this bipartite graph (bipartite because we have a set of control
vertices that are only connected to target vertices, and a set of
target vertices that are only connected to control vertices). By
Hall's theorem \cite{steane:fast}, there is a coloring using $w$
colors, and $w$ colors is the minimum number of colors we can use.
Here $w$ is maximum weight of the rows of $A$ minus 1. See
\cite{KR:edgecolor} for an algorithm that finds an edge coloring
with $w$ colors in time $O(n N_{{\rm CNOT}})$. Here $n$ is the
number of qubits that are to be encoded (i.e. number of vertices)
and $N_{\rm CNOT}$ is the number of CNOT gates in the encoder
(i.e. number of edges).


We have tested several of the encoders produced by our Latin
rectangle software tool with and without memory locations by
simulating them using CHP\footnote{CHP stands for CNOT, Hadamard, Phase.} \cite{AG:stabilizer}.

\section{Syndrome Decoding}
\label{app:decode}

General algorithms for constructing the classical circuits to
decode measurement outcomes obtained in Steane error correction
require exponential time and/or space. Therefore, we consider each
code's syndrome decoder separately, essentially devising a
special-purpose algorithm for each to make the decoding feasible.

\begin{table}
\begin{center}
\begin{tabular}{l|l|l}
\textsc{Code}  & \textsc{Decoder}
\\ \hline
$[[5,1,3]]$ & Table Lookup\\
$[[7,1,3]]$ & Table Lookup (cyclic)\\
$[[9,1,3]],[[25,1,5]],[[49,1,7]],[[81,1,9]]$ & Majority \\
$[[15,1,3]]$ & Table Lookup \\
$[[13,1,3]],[[41,1,5]],[[85,1,7]]$ & Min. Wt. Matching \\
$[[21,3,5]]$ & Table Lookup (cyclic)\\
$[[23,1,7]]$ & Table Lookup (cyclic)\\
$[[47,1,11]]$ & Algebraic \cite{chen:decodingQR} \\
$[[49,1,9]]$ & Table Lookup with Message Passing \\
$[[60,4,10]]$ & Table Lookup (cyclic)\\
\end{tabular}
\end{center}
\caption{The decoders that we use for the codes in our study.}
\label{tab:decoders}
\end{table}

Table~\ref{tab:decoders} lists all of the codes we consider in this
study and their syndrome decoders. There are six distinct decoding
algorithms that we use to compute the error locations and type of
error from the syndrome measurements: a generic table lookup algorithm,
a table lookup algorithm for cyclic codes over arbitrary fields, a majority voting algorithm for
Bacon-Shor codes, a minimum weight matching algorithm for surface
codes, a simple message passing algorithm for the concatenated
Hamming code, and an algebraic decoder for the $[[47,1,11]]$
quadratic residue code.

Rather than use a general table-lookup algorithm, we use a so-called
Meggitt decoder which uses the fact that the polynomial codes and
the Hamming, Golay, and quadratic residue (QR) codes are constructed
from cyclic classical codes. Cyclic codes have a compact description
in terms of a generating polynomial whose coefficients give one of
the code words and whose cyclic shifts generate a basis for the
code. The Meggitt decoding algorithm stores a table of syndromes and
their associated error corrections \cite{book:HP}. For non-binary
codes such as the polynomial codes, the table stores both error
locations and error-type (the so-called amplitude).
Only $n\choose w-1$ syndromes need to be stored for a weight $w$
error, since one of the coordinates can be fixed by the cyclic
symmetry. Finding the appropriate recovery requires at most $n$
table lookups. If we fail to find a recovery in the table, a subroutine
is triggered that applies some syndrome-dependent correction mapping the
state back into the code space.

For cyclic codes with larger distance where table lookup is
impractical, for example $[[47,1,11]]$, algebraic decoding
techniques can be used. The generator polynomial's roots are used
to compute a sequence of syndromes from which we can locate
errors. BCH codes are easy to decode because their generator
polynomials have a contiguous sequence of roots so the
Berlekamp-Massey algorithm can find the error-locator polynomial
whose roots give the error locations. Sometimes decoding up to the
full minimum distance of the code is challenging because the
generator polynomial may not have a long sequence of roots, so
some syndromes are missing and the Berlekamp-Massey algorithm
cannot be directly applied. In this case, unknown syndromes can
sometimes be computed from algebraic equations involving the known
syndromes. Algebraic decoding of the $[[47,1,11]]$ proceeds this
way. For each error weight from zero to $t$, we compute any
missing syndromes, construct a polynomial whose roots are the
error locations, and find the roots of the polynomial. If the
polynomial has enough roots, we correct those errors and stop. If
we do not find enough roots for each of the locators, we return a
``failed'' result, triggering a subroutine that applies some
syndrome-dependent correction that maps the state to some
(possibly logically incorrect) state in the code space. The
implementation details can be found in \cite{chen:decodingQR}.

The Bacon-Shor codes are essentially concatenated quantum repetition
codes. Since the code stabilizer is preserved by bitwise Hadamard
composed with a $90$ degree rotation of the square lattice, one
syndrome decoder is sufficient for both $X$ and $Z$ error
correction. Imagine a vector of $n^2$ syndrome bits placed on an $n$
by $n$ square lattice. Let $s_x$ be the syndrome vector for $X$
errors and $s_z$ be the syndrome vector for $Z$ errors. Let $R$ be
the map on vectors of length $n^2$ that rotates them by $90$ degrees
on the square lattice. The same syndrome decoder is applied to $s_x$
and $Rs_z$. The syndrome decoder decodes a variation on the
classical repetition code on $n$ bits. First, the decoder computes
the parity of each column of the lattice and stores each column
parity as an element of a vector $p$. Next, the decoder computes the
repetition code parity check $h=Hp$. This parity check $H$ is
expressed in standard form $[I_{n-1}\ 1]$ where $1$ is the all ones
column vector. Finally, the decoder infers the error locations from
the parity check. If the weight of the parity check is greater than
$t$, we must assume that the rightmost bit of $p$ was incorrect so
that $h\oplus 1$ gives the error locations on the first $n-1$ bits
of $p$. Otherwise, we infer that the rightmost bit of $p$ was
correct so that $h$ gives the error locations on the first $n-1$
bits of $p$.

The surface code is decoded using Edmond's minimum-weight matching
algorithm. The approach differs slightly depending on
whether Steane-EC or Shor-EC is used but is essentially the same as \cite{dennis+:top}.
Steane-EC gives a 2D matching problem whereas Shor-EC gives a 3D matching problem.
The mapping from syndrome information to a matching problem is as follows.


Nonzero syndrome bits are called defects and are located somewhere
in the $\ell \times \ell$ plane. We construct a complete weighted
graph whose vertices represent defects and whose edge weights
indicate the distance between defects. The surface code's syndrome
may be such that there are lone defects which are not caused by
error patterns connecting two defects, but by an error pattern
connecting an edge-defect on the boundary to an inner defect. $X$
and $Z$ errors constitute separate matching problems and $X$-defects
can be matched with, say, the horizontal boundaries and $Z$-defects
with the vertical boundaries.


We can design an algorithm for decoding the surface code for, say,
$Z$ errors, as follows:
\begin{itemize}
\item Imagine cutting the lattice vertically in two halves, left (L) and right (R).
Let $N_{L/R}(i)$ be the number of defects in row $i$ of the
left/right part of the lattice.  For each row of the lattice, add
$N_{L/R}(i)$ edge defects on the $i$th row on the left (right)
boundary.
\item Assign the weight of the edges between {\em any} edge defects as zero and assign
the distance as the weight between edge defects and inner defects.
\item Compute the minimum-weight perfect matching of the graph of defects.
\item The recovery operation consists of applying phase flips on the qubits that are
along the edges of each pair of matched vertices in the graph.
\end{itemize}

Note that the algorithm enforces the property that the graph has an
even number of vertices, so that every vertex can be matched.


\begin{figure}[htb]
\centerline{ \mbox{ \Qcircuit @C=1em @R=.5em {
\lstick{\ket{+}} & \ctrl{2} & \ctrl{4} & \qw       & \qw             & \gate{{\rm S}}      & \gate{{\rm Cyc}^\dag} & \gate{Z}        & \multigate{4}{ED_z} & \multigate{9}{M_{\overline{X}\overline{X}}} & \multigate{4}{ED} & \multigate{9}{M_{\overline{X}\overline{X}}} & \qw \\
\lstick{\ket{0}} & \qw      & \qw      & \targ     & \qw             & \qw           & \targ           & \gate{{\rm Cyc}^\dag} & \ghost{ED_z}        & \ghost{\cal{M}_{\overline{X}\overline{X}}} & \ghost{ED} & \ghost{\cal{M}_{\overline{X}\overline{X}}} & \qw \\
\lstick{\ket{0}} & \targ    & \qw      & \qw       & \gate{{\rm Cyc}^\dag} & \gate{{\rm S}^\dag} & \ctrl{-1}       & \gate{{\rm S}}        & \ghost{ED_z}        & \ghost{\cal{M}_{\overline{X}\overline{X}}} & \ghost{ED} & \ghost{\cal{M}_{\overline{X}\overline{X}}} & \qw \\
\lstick{\ket{+}} & \ctrl{1} & \qw      & \ctrl{-2} & \qw             & \gate{{\rm S}}      & \gate{{\rm Cyc}^\dag} & \gate{Z}        & \ghost{ED_z}        & \ghost{\cal{M}_{\overline{X}\overline{X}}} & \ghost{ED} & \ghost{\cal{M}_{\overline{X}\overline{X}}} & \qw \\
\lstick{\ket{0}} & \targ    & \targ    & \qw       & \qw             & \gate{X}      & \qw             & \qw             & \ghost{ED_z}        & \ghost{\cal{M}_{\overline{X}\overline{X}}} & \ghost{ED} & \ghost{\cal{M}_{\overline{X}\overline{X}}} & \qw \\
\lstick{\ket{+}} & \ctrl{2} & \ctrl{4} & \qw       & \qw             & \gate{{\rm S}}      & \gate{{\rm Cyc}^\dag} & \gate{Z}        & \multigate{4}{ED_z} & \ghost{\cal{M}_{\overline{X}\overline{X}}} & \multigate{4}{ED} & \ghost{\cal{M}_{\overline{X}\overline{X}}} & \qw \\
\lstick{\ket{0}} & \qw      & \qw      & \targ     & \qw             & \qw           & \targ           & \gate{{\rm Cyc}^\dag} & \ghost{ED_z}        & \ghost{\cal{M}_{\overline{X}\overline{X}}} & \ghost{ED} & \ghost{\cal{M}_{\overline{X}\overline{X}}} & \qw \\
\lstick{\ket{0}} & \targ    & \qw      & \qw       & \gate{{\rm Cyc}^\dag} & \gate{{\rm S}^\dag} & \ctrl{-1}       & \gate{{\rm S}}        & \ghost{ED_z}        & \ghost{\cal{M}_{\overline{X}\overline{X}}} & \ghost{ED} & \ghost{\cal{M}_{\overline{X}\overline{X}}} & \qw \\
\lstick{\ket{+}} & \ctrl{1} & \qw      & \ctrl{-2} & \qw             & \gate{{\rm S}}      & \gate{{\rm Cyc}^\dag} & \gate{Z}        & \ghost{ED_z}        & \ghost{\cal{M}_{\overline{X}\overline{X}}} & \ghost{ED} & \ghost{\cal{M}_{\overline{X}\overline{X}}} & \qw \\
\lstick{\ket{0}} & \targ    & \targ    & \qw       & \qw &
\gate{X} & \qw             & \qw             & \ghost{ED_z} &
\ghost{\cal{M}_{\overline{X}\overline{X}}} & \ghost{ED} &
\ghost{\cal{M}_{\overline{X}\overline{X}}} & \qw }}} \caption{A
fault-tolerant circuit for preparing logical Bell pairs for Knill
error correction of $[[5,1,3]]$. The sub-circuit $ED_z$ measures
the stabilizer of $\ket{\overline{0}}$, $\langle XZZXI, IXZZX,
XIXZZ, ZXIXZ, ZZZZZ\rangle$, using $4$ and $5$ qubit cat states,
and the sub-circuit $ED$ makes the same measurement without
measuring $\overline{Z}=ZZZZZ$. If any measurement outcome is
nonzero, the Bell state is rejected. The sub-circuit
$M_{\overline{X}\overline{X}}$ measures
$\overline{X}\overline{X}=X^{\otimes 10}$ using a $10$ qubit cat
state. The Bell state is rejected if the
$\overline{X}\overline{X}$ measurements disagree, but if they are
both $1$ then $\overline{Z}_1$ is applied to the output Bell pair.
The cat states are verified so that if a cat state is accepted
then a single fault in its preparation cannot produce a correlated
error.} \label{fig:bell513}
\end{figure}

The concatenated [[7,1,3]] code, that is, the [[49,1,9]] code, can
be decoded to distance $7$ if we treat it as a concatenated code.
However, decoding the code to distance $9$ requires a slight
modification of the algorithm so that a simple message is passed
from level-1 to level-2.

Suppose the $49$ transversal measurement outcomes are organized into
$7$ registers of $7$ bits each. We use these registers as temporary storage
to compute the appropriate correction. First, we compute the level-1 syndromes for
each register as we would normally do. These syndromes indicate errors $e_i$
in the $i$th level-1 register. We correct each level-1 register according
to the $e_i$s and ``flag'' those registers for which $e_i\neq 0$.
Next, we compute the level-2 (logical) syndrome of the resulting $49$ bit
register, which now has trivial level-1 syndrome in each $7$ bit register.
This level-2 syndrome indicates a logical correction $\bar{e}$ that is
constant on each level-1 register (but two level-1 registers can take
different values).
The correction $c_1:=\left(\bigoplus_i e_i\right)\oplus\bar{e}$ corrects
all errors of weight $4$ or less, except for one problem case. This case
occurs when a pair of errors occurs in one level-1 register and another
pair of errors occurs in a different level-1 register. The problem is
overcome by comparing the register positions where $\bar{e}$ is $1$ with the
positions of the flags whenever two flags are raised. If they disagree,
apply the correction $c_2:=\left(\bigoplus_i e_i\right)\oplus\bar{f}$
where $\bar{f}$ is a logical correction on the flagged registers. Otherwise,
apply the original correction $c_1$. This procedure corrects all errors
of weight $4$ or less and returns the input to the codespace in all cases.

The classical decoding algorithms have been tested exhaustively for all
of the codes in the paper except for the large Bacon-Shor and surface codes.
The algorithms were found to correct all errors of weight $t$ or less. In
the case of the polynomial codes, $t$ is the number of errors the underlying
nonbinary code can correct. The same decoding algorithms that were tested
exhaustively for small Bacon-Shor and surface codes were used for the larger
codes in those families.

\section{Data Tables}
\label{app:data}


\pagestyle{empty}
\setlength\topmargin{0.5in} \setlength\headheight{0in}
\setlength\headsep{0in} \setlength\textheight{10in}
\setlength\textwidth{6.5in} \setlength\oddsidemargin{0in}
\setlength\evensidemargin{0in} \setlength\parindent{0.0in}
\setlength\parskip{0.25in}

\begin{landscape}
\begin{table}[htp]
\centering
\begin{minipage}{\textwidth}
\centering
\begin{tabular}{l|l|l|l|l|l|l|l}
$[[n,k,d]]$ & \textsc{L}\footnote{for $[[5,1,3]]$ this parameter is \textsc{NB}.} & \textsc{R}\footnote{for $[[5,1,3]]$ this parameter is \textsc{NC}.} &
\textsc{CX/Rec}\footnote{for $[[5,1,3]]$ this parameter is the number of CNOT gates in a $T_3$ rectangle} & $p_1(p_{\rm mem}=0,p_0=10^{-4})$ & $p_1(p_{\rm mem}=p_0=10^{-4})$ &
$p_{\rm th}(p_{\rm mem}=0)$ & $p_{\rm th}(p_{\rm mem}=p_0)$ \\
\hline\hline
$[[5,1,3]]$ & 2  &    2    & 2,160 & -- & -- & \dat{3.9}{0.7}{5} & \dat{2.5}{0.4}{5}       \\
$[[5,1,3]]$ & 3  &    3    & 5,117 & -- & -- & \dat{9.2}{0.5}{5} & \dat{3.7}{0.3}{5}       \\
$[[5,1,3]]$ & 5  &    5    & 14,775 & -- & -- & \dat{9.2}{0.5}{5} & \dat{3.3}{0.6}{5}       \\
$[[5,1,3]]$ & 10 &    3    & 18,536 & -- & -- & \dat{8.8}{0.5}{5} & \dat{4.3}{0.3}{5}       \\
$[[5,1,3]]$ & 10 &    10   & 60,760 & -- & -- & \dat{8}{3}{5}     & \dat{3.0}{0.6}{5}        \\ \hline
$[[7,1,3]]$ & 2  &     1   & 519      &       \dat{5.34}{0.07}{4} & \dat{7.05}{0.08}{4} &   \dat{1.85}{0.05}{5}  &  \dat{1.46}{0.05}{5} \\
$[[7,1,3]]$ & 3  &     1   & 775      &       \dat{2.3}{0.2}{5}  & \dat{4.5}{0.2}{5}  &   \dat{3.11}{0.02}{4} &  \dat{1.98}{0.01}{4}  \\
$[[7,1,3]]$ & 4  &     1   & 1,031     &       \dat{1.9}{0.1}{5}  & \dat{3.7}{0.2}{5}  &   \dat{4.97}{0.07}{4} &  \dat{2.56}{0.06}{4}  \\
$[[7,1,3]]$ & 5  &     1   & 1,287     &       \dat{1.8}{0.1}{5}  & \dat{4.1}{0.2}{5}  &   \dat{5.3}{0.1}{4}   & \dat{2.58}{0.06}{4} \\ \hline
$[[9,1,3]]$ & 1  &     1   & 69       &       -- & \dat{4.90}{0.09}{5} & \dat{2.6}{0.1}{4} & \dat{2.06}{0.02}{4} \\ \hline
$[[13,1,3]]$ & 3   &    1   &  1,501     &       --               &  --   & \dat{1.59}{0.04}{4} & \dat{0.69}{0.03}{4} \\
$[[13,1,3]]$ & 4   &    1   &  1,997     &      --               &  --   & \dat{3.81}{0.07}{4} & \dat{1.95}{0.04}{4} \\
$[[13,1,3]]$ & 5   &    1   &  2,493     &      --               &   \dat{4.3}{0.2}{5}  & \dat{4.9}{0.2}{4}   & \dat{2.30}{0.08}{4} \\
$[[13,1,3]]$ & 10  &    1   &  4,973     &      \dat{1.8}{0.1}{5} &  \dat{3.9}{0.2}{5}  & \dat{5.1}{0.2}{4}   & \dat{2.54}{0.07}{4} \\
$[[13,1,3]]$ & 15  &    1   &  7,453     & \dat{1.9}{0.1}{5} &  \dat{3.9}{0.2}{5}  & \dat{4.9}{0.1}{4}   & \dat{2.63}{0.07}{4} \\ \hline
$[[15,1,3]]$ &  3  &    1   & 2,127  &  \dat{1.3}{0.2}{4}    & \dat{4.5}{0.2}{4} & \dat{0.86}{0.03}{4}  & \dat{0.33}{0.05}{4} \\
$[[15,1,3]]$ &  4  &    1   & 2,831  &  \dat{4.9}{0.7}{5} & \dat{1.0}{0.1}{4} & \dat{1.5}{0.6}{4}    & \dat{1.0}{0.2}{4} \\
$[[15,1,3]]$ &  5  &    1   & 3,535  & \dat{5.8}{0.8}{5} & \dat{1.0}{0.1}{4} & \dat{1.8}{0.2}{4}    & \dat{1.0}{0.2}{4} \\ \hline
$[[23,1,7]]$ & 10 & 1   &   16,023 & \dat{1.1}{0.3}{7} & \dat{1.2}{0.6}{7} & \dat{1.14}{0.05}{3} & \dat{1.09}{0.01}{3} \\
$[[23,1,7]]$ & 20 & 1   &   32,023 & \dat{1.2}{0.4}{7} & \dat{9}{4}{8} & \dat{2.33}{0.02}{3} & \dat{1.97}{0.02}{3} \\
$[[23,1,7]]$ & 30 & 1   &   48,023 & --                & -- & \dat{2.98}{0.04}{3} & \dat{2.25}{0.03}{3} \\
$[[23,1,7]]$ & 40 & 1   &   64,023 & --                & -- & \dat{3.33}{0.02}{3} & \dat{2.19}{0.04}{3} \\
$[[23,1,7]]$ & 10 & 2   &   28,023  & \dat{4}{1}{8} & \dat{3}{2}{8} & \dat{5.76}{0.09}{4} & \dat{5.48}{0.09}{4} \\
$[[23,1,7]]$ & 20 & 2   &   56,023  & \dat{5}{1}{8} & $\approx <4\times 10^{-8}$ \footnote{one failure in $5\times 10^{7}$ samples} & \dat{1.23}{0.01}{3} & \dat{1.15}{0.01}{3} \\
$[[23,1,7]]$ & 30 & 2   &   84,023  & --              & -- & \dat{1.628}{0.006}{3} & \dat{1.487}{0.003}{3} \\
$[[23,1,7]]$ & 40 & 2   &   112,023 & --              & -- & \dat{1.95}{0.01}{3} & \dat{1.77}{0.02}{3} \\
\end{tabular}
\end{minipage}
\caption{Complete tabulation of code survey data, part 1}
\label{table:complete1}
\end{table}
\end{landscape}

\begin{landscape}
\begin{table}[htp]
\centering
\begin{minipage}{\textwidth}
\centering
\begin{tabular}{l|l|l|l|l|l|l|l}
$[[n,k,d]]$ & \textsc{L} & \textsc{R}\footnote{for $[[49,1,9]]$ this parameter is the number of preparation attempts for a 7-qubit encoded ancilla used in error detection} & \textsc{CX/Rec} & $p_1(p_{\rm mem}=0,p_0=10^{-4})$ & $p_1(p_{\rm mem}=p_0=10^{-4})$ & $p_{\rm th}(p_{\rm mem}=0)$ & $p_{\rm th}(p_{\rm mem}=p_0)$ \\
\hline\hline
$[[23,1,7]]$ & 10 &     3   &   40,023   & \dat{4}{2}{8} & \dat{3}{1}{8} & \dat{3.72}{0.05}{4} & \dat{3.45}{0.05}{4} \\
$[[23,1,7]]$ & 20 &     3   &   80,023   & --            & -- & \dat{8.03}{0.05}{4} & \dat{7.67}{0.05}{4} \\
$[[23,1,7]]$ & 30 &     3   &   120,023  & --            & -- & \dat{1.095}{0.003}{3} & \dat{1.036}{0.008}{3} \\
$[[23,1,7]]$ & 40 &     3   &   160,023  & -- & -- & \dat{1.366}{0.007}{3} & \dat{1.280}{0.009}{3} \\ \hline
$[[25,1,5]]$ & 4  & 1   & 1,465    & -- & \dat{1.2}{0.7}{6} & \dat{8.6}{0.2}{4} & \dat{7.44}{0.05}{4} \\
$[[25,1,5]]$ & 5  & 1   & 1,825    & -- & \dat{1.0}{0.1}{6} & \dat{1.13}{0.02}{3} & \dat{9.74}{0.07}{4} \\
$[[25,1,5]]$ & 6  & 1   & 2,185    & -- & \dat{1.08}{0.08}{6}  & \dat{1.16}{0.02}{3} & \dat{1.034}{0.008}{3} \\
$[[25,1,5]]$ & 7  & 1   & 2,545    & -- & \dat{1.1}{0.2}{6} & \dat{1.17}{0.04}{3} & \dat{1.01}{0.04}{3} \\ \hline
$[[41,1,5]]$ & 5  & 1   & 11,321    & \dat{7.3}{0.8}{6} & \dat{2.39}{0.05}{4} & \dat{1.86}{0.02}{4} & \dat{7.9}{0.1}{5} \\
$[[41,1,5]]$ & 10 & 1   & 22,601    & $[3\times 10^{-7}]$\footnote{The values in square brackets are extrapolated from a linear least-squares fit to the logarithm of $p_1(p_0)$}  & $[7\times 10^{-7}]$ & \dat{7.44}{0.03}{4} & \dat{3.44}{0.01}{4} \\
$[[41,1,5]]$ & 15 & 1   & 33,881    & --                & -- & \dat{1.224}{0.003}{3} & \dat{5.55}{0.02}{4} \\
$[[41,1,5]]$ & 20 & 1   & 45,161    & --                & -- & \dat{1.577}{0.004}{3} & \dat{7.61}{0.02}{4} \\
$[[41,1,5]]$ & 30 & 1   & 67,721    & -- & -- & \dat{2.06}{0.01}{3} & \dat{1.008}{0.008}{3} \\ \hline
$[[47,1,11]]$ & 10 &    1 & 52,527  &  -- & -- & \dat{3.25}{0.04}{4} & \dat{2.15}{0.04}{4} \\
$[[47,1,11]]$ & 20 &    1 & 105,007 &  -- & -- & \dat{6.89}{0.05}{4} & \dat{4.79}{0.03}{4} \\
$[[47,1,11]]$ & 30 &    1 & 157,487 & \dat{1.6}{0.9}{7} & -- & \dat{9.51}{0.04}{4} & \dat{6.45}{0.03}{4} \\ \hline
$[[49,1,9]]$ & 5  & 2 & 61,549   & -- & -- & \dat{1.02}{0.02}{4} & \dat{5.4}{0.1}{5} \\
$[[49,1,9]]$ & 10 & 2 & 123,049  & -- & -- & \dat{3.63}{0.08}{4} & \dat{2.23}{0.04}{4} \\
$[[49,1,9]]$ & 15 & 2 & 184,549  & -- & -- & \dat{4.0}{0.02}{4} & \dat{3.20}{0.08}{4} \\
$[[49,1,9]]$ & 20 & 2 & 246,049  & \dat{4}{1}{6} & -- & \dat{4.2}{0.3}{4} & -- \\ \hline
$[[49,1,7]]$ & 4  & 1 & 2,961 & -- & \dat{3.4}{0.2}{6} & \dat{4.73}{0.09}{4} & \dat{3.20}{0.02}{4} \\
$[[49,1,7]]$ & 6  & 1 & 4,417 & -- & \dat{2.8}{0.2}{7} & \dat{1.18}{0.01}{3} & \dat{8.7}{0.2}{4} \\
$[[49,1,7]]$ & 8  & 1 & 5,873 & -- & \dat{3.3}{0.6}{7} & \dat{1.41}{0.02}{3} & \dat{1.169}{0.005}{3} \\
$[[49,1,7]]$ & 9  & 1 & 6,601 & -- & \dat{2.2}{0.7}{7} & \dat{1.48}{0.02}{3} & \dat{1.224}{0.005}{3} \\
$[[49,1,7]]$ & 10 & 1 & 7,329 & -- & \dat{4.0}{0.9}{7} & \dat{1.42}{0.03}{3} & \dat{1.235}{0.005}{3} \\
$[[49,1,7]]$ & 11 & 1 & 8,057 & -- & \dat{2.5}{0.2}{7} & \dat{1.46}{0.03}{3} & \dat{1.241}{0.006}{3} \\
$[[49,1,7]]$ & 12 & 1 & 8,785 & -- & \dat{3}{2}{7} & \dat{1.46}{0.02}{3} & \dat{1.242}{0.006}{3} \\ \hline
$[[60,4,10]]$ & 10 &    1 & 86,460  & -- & -- & \dat{1.129}{0.004}{4} & -- \\
$[[60,4,10]]$ & 20 &    1 & 172,860 & -- & -- & \dat{3.91}{0.02}{4} & \dat{2.20}{0.04}{4}
\end{tabular}
\end{minipage}
\caption{Complete tabulation of code survey data, part 2}
\label{table:complete2}
\end{table}
\end{landscape}

\begin{landscape}
\begin{table}[htp]
\centering
\begin{minipage}{\textwidth}
\centering
\begin{tabular}{l|l|l|l|l|l|l|l}
$[[n,k,d]]$ & \textsc{L} & \textsc{R} & \textsc{CX/Rec} & $p_1(p_{\rm mem}=0,p_0=10^{-4})$ & $p_1(p_{\rm mem}=p_0=10^{-4})$ & $p_{\rm th}(p_{\rm mem}=0)$ & $p_{\rm th}(p_{\rm mem}=p_0)$ \\
\hline\hline
$[[81,1,9]]$ & 4  & 1 & 4,977   &  -- & \dat{4.4}{0.7}{5}   & \dat{2.1}{0.2}{4} & \dat{1.407}{0.005}{4} \\
$[[81,1,9]]$ & 6  & 1 & 7,425   &  -- & $[1\times 10^{-6}]$\footnote{The values in square brackets are extrapolated from a linear least-squares fit to the logarithm of $p_1(p_0)$}   & \dat{7.1}{0.1}{4} & \dat{4.47}{0.03}{4} \\
$[[81,1,9]]$ & 10 & 1 & 12,321  &  -- & $[7\times 10^{-7}]$   & \dat{1.25}{0.02}{3} & \dat{9.57}{0.03}{4} \\
$[[81,1,9]]$ & 11 & 1 & 13,545  &  -- & --   & \dat{1.32}{0.02}{3} & \dat{1.029}{0.004}{3} \\
$[[81,1,9]]$ & 12 & 1 & 14,769  &  -- & --   & \dat{1.29}{0.03}{3} & \dat{1.069}{0.006}{3} \\
$[[81,1,9]]$ & 18 & 1 & 22,113  &  -- & --   & \dat{1.30}{0.03}{3} & \dat{1.113}{0.006}{3} \\
$[[81,1,9]]$ & 19 & 1 & 23,337  &  -- & --   & \dat{1.34}{0.03}{3} & \dat{1.098}{0.006}{3} \\
$[[81,1,9]]$ & 20 & 1 & 24,561  &  -- & --   & \dat{1.34}{0.02}{3} & \dat{1.112}{0.006}{3} \\ \hline
$[[85,1,7]]$ & 5  &    1 & 30,405  &  -- &   -- & \dat{5.7}{0.1}{5} & \dat{2.03}{0.07}{5} \\
$[[85,1,7]]$ & 10 &    1 & 60,725  &  -- &   -- & \dat{2.48}{0.01}{4} & \dat{1.03}{0.04}{4} \\
$[[85,1,7]]$ & 15 &    1 & 91,045  &  -- &   -- & \dat{4.18}{0.05}{4} & \dat{1.76}{0.02}{4} \\
$[[85,1,7]]$ & 20 &    1 & 121,365 &  -- & $[2\times 10^{-7}]$  & \dat{5.59}{0.04}{4} & \dat{2.32}{0.02}{4}
\end{tabular}
\end{minipage}
\caption{Complete tabulation of code survey data, part 3}
\label{table:complete3}
\end{table}
\end{landscape}

\begin{landscape}
\begin{table}[htp]
\centering
\begin{minipage}{\textwidth}
\centering
\begin{tabular}{l|l|l|l}
$[[n,k,d=\ell]]$ & \textsc{CX/Rec} & $p_1(p_{\rm mem}=p_0=10^{-4})$ & $p_{\rm th}(p_{\rm mem}=p_0)$ \\
\hline\hline
$[[41,1,5]]$   & 1,481  & \dat{1.7}{0.1}{4}   & \dat{6.8}{0.6}{5} \\
$[[85,1,7]]$   & 4,453  & \dat{5}{2}{5}       & \dat{2.3}{0.2}{4} \\
$[[145,1,9]]$  & 9,937  & \dat{2}{1}{5}       & \dat{4.5}{0.2}{4} \\
$[[221,1,11]]$ & 18,701 & $[8\times 10^{-6}]$\footnote{The values in square brackets are extrapolated from a linear least-squares fit to the logarithm of $p_1(p_0)$} & \dat{6.6}{0.2}{4} \\
$[[313,1,13]]$ & 31,513 & $[8\times 10^{-6}]$ & \dat{9.0}{0.4}{4}
\end{tabular}
\end{minipage}
\caption{Surface code data using Shor-EC and a transversal CNOT as in \cite{dennis+:top}, taking $\ell$ syndromes for an $\ell\times\ell$ code EC.}
\label{table:surfacedat}
\end{table}
\end{landscape}

\begin{landscape}
\begin{table}[htp]
\centering
\begin{minipage}{\textwidth}
\centering
\begin{tabular}{l|l|l}
$[[n,k,d]]$     & family               & $p_{\rm th}(\textrm{perfect ancilla})$ \\ \hline\hline
$[[5,1,3]]$     &  &    \dat{2.0}{0.1}{4} \\
$[[7,1,3]]$ & doubly-even dual-containing &     \dat{9.1}{0.2}{4} \\
$[[9,1,3]]$ & Bacon-Shor     &  \dat{6.0}{0.9}{4} \\
$[[13,1,3]]$    & surface &     \dat{8.8}{0.1}{4} \\
$[[21,3,5]]$    & polynomial &  $<10^{-5}$ \\
$[[23,1,7]]$    & dual-containing & \dat{5.34}{0.04}{3} \\
$[[25,1,5]]$    & Bacon-Shor &  \dat{1.88}{0.04}{3} \\
$[[41,1,5]]$    & surface &     \dat{3.8}{0.3}{3} \\
$[[47,1,11]$    & doubly-even dual-containing & \dat{7.67}{0.03}{3} \\
$[[49,1,7]]$    & Bacon-Shor &  \dat{2.56}{0.05}{3} \\
$[[49,1,9]]$    & doubly-even dual-containing & \dat{4.8}{0.2}{3} \\
$[[60,4,10]]$   & polynomial &  \dat{1.88}{0.04}{3} \\
$[[81,1,9]]$    & Bacon-Shor &  \dat{2.88}{0.04}{3} \\
$[[85,1,7]]$    & surface &     \dat{7.5}{0.3}{3} \\
$[[121,1,11]]$  & Bacon-Shor &  \dat{2.83}{0.07}{3} \\
$[[145,1,9]]$   & surface &     \dat{1.01}{0.02}{2} \\
$[[169,1,13]]$  & Bacon-Shor &  \dat{2.97}{0.09}{3}
\end{tabular}
\end{minipage}
\caption{Level-1 pseudo-thresholds for rectangles using Steane-EC with perfect (noiseless) ancilla, $n<200$.}
\label{table:perfectdat}
\end{table}
\end{landscape}




\end{document}